\documentclass[aps,twocolumn,superscriptaddress,floatfix,preprintnumbers]{revtex4-2}
\usepackage[utf8]{inputenc}  
\usepackage{graphicx}
\usepackage{dcolumn}
\usepackage{bm}
\usepackage{color,soul,xcolor}
\usepackage{amsmath}
\usepackage{booktabs}
\usepackage{flushend}
\usepackage{xr}
\usepackage{placeins}
\usepackage{subfigure}
\usepackage{subcaption}
\usepackage[pdftex,bookmarks,linktocpage,pdfpagelabels,plainpages=false,hyperfigures,linkcolor=blue,citecolor=blue]{hyperref} 
\usepackage[singlelinecheck=false,justification=raggedright]{caption}

\newlength{\wdth}

\usepackage{ulem,lipsum}
\newcommand\xoutpars[1]{\let\helpcmd\xout\parhelp#1\par\relax\relax}
\newcommand\soutpars[1]{\let\helpcmd\sout\parhelp#1\par\relax\relax}
\long\def\parhelp#1\par#2\relax{%
  \helpcmd{#1}\ifx\relax#2\else\par\parhelp#2\relax\fi%
}

\begin{document}
\begin{flushright}
FERMILAB-PUB-24-0873
\end{flushright}

\title{Search for Inelastic Boosted Dark Matter with the ICARUS Detector \\
at the Gran Sasso Underground National Laboratory}

\author{H. Carranza}
\affiliation{Department of Physics, The University of Texas, Arlington, TX76019, USA}
\author{J. Yu}
\affiliation{Department of Physics, The University of Texas, Arlington, TX76019, USA}
\author{B. Brown}
\affiliation{Department of Physics, The University of Texas, Arlington, TX76019, USA}
\author{S. Blanchard}
\affiliation{Department of Physics, The University of Texas, Arlington, TX76019, USA}
\author{S. Chakraborty}
\affiliation{Department of Physics, The University of Texas, Arlington, TX76019, USA}
\author{R. Raut}
\affiliation{Department of Physics, The University of Texas, Arlington, TX76019, USA}

\collaboration{Department of Physics, The University of Texas at Arlington}

\author{D. Kim}
\affiliation{Department of Physics, The University of South Dakota, Vermillion, SD57069, USA}

\collaboration{Department of Physics, The University of South Dakota}

\author{M. Antonello}
\affiliation{INFN - Laboratori Nazionali del Gran Sasso, Assergi, Italy}
\author{B. Baibussinov}
\affiliation{Dipartimento di Fisica e Astronomia Universit\`{a} di Padova and INFN, Padova, Italy}
\author{V. Bellini}
\affiliation{Dipartimento di Fisica e Astronomia Universit\`{a} di Catania and INFN, Catania, Italy}
\author{P. Benetti}
\affiliation{Dipartimento di Fisica Universit\`{a} di Pavia and INFN, Pavia, Italy}
\author{F.Boffelli}
\affiliation{Dipartimento di Fisica Universit\`{a} di Pavia and INFN, Pavia, Italy}
\author{M. Bonesini}
\affiliation{Dipartimento di Fisica Universit\`{a} di Milano Bicocca and INFN, Milano, Italy}
\author{A. Bubak}
\affiliation{ Institute of Physics, University of Silesia, Katowice, Poland}
\author{E. Calligarich}
\affiliation{Dipartimento di Fisica Universit\`{a} di Pavia and INFN, Pavia, Italy}
\author{S.Centro}
\affiliation{Dipartimento di Fisica e Astronomia Universit\`{a} di Padova and INFN, Padova, Italy}
\author{A. Cesana}
\affiliation{Dipartimento di Fisica Universit\`{a} di Milano and INFN, Milano, Italy}
\author{K. Cieslik}
\affiliation{H. Niewodniczanski Institute of Nuclear Physics, Polish Academy of Science, Krakow, Poland}
\author{A.G. Cocco}
\affiliation{INFN - Laboratori Nazionali del Gran Sasso, Assergi, Italy}
\author{A. Dabrowska}
\affiliation{H. Niewodniczanski Institute of Nuclear Physics, Polish Academy of Science, Krakow, Poland}
\author{A. Dermenev}
\affiliation{INR RAS, Moscow, Russia}
\author{A. Falcone}
\affiliation{Dipartimento di Fisica Universit\`{a} di Milano Bicocca and INFN, Milano, Italy}
\author{C. Farnese}
\affiliation{Dipartimento di Fisica e Astronomia Universit\`{a} di Padova and INFN, Padova, Italy}
\author{A. Fava}
\affiliation{Fermi National Accelerator Laboratory, Batavia, IL 60510, USA}
\author{A. Ferrari}
\affiliation{Institute of Astroparticle Physics, Karlsruhe Institute of Technology, Karlsruhe (KIT), Germany}
\affiliation{Institute for Beam Physics and Technology, Karlsruhe Institute of Technology (KIT), Karlsruhe, Germany} 
\author{D. Gibin}
\affiliation{Dipartimento di Fisica e Astronomia Universit\`{a} di Padova and INFN, Padova, Italy}
\author{S. Gninenko}
\affiliation{INR RAS, Moscow, Russia}
\author{A. Guglielmi}
\affiliation{Dipartimento di Fisica e Astronomia Universit\`{a} di Padova and INFN, Padova, Italy}
\author{J. Holeczek}
\affiliation{Institute of Physics, University of Silesia, Katowice, Poland}
\author{M. Janik}
\affiliation{Institute of Physics, University of Silesia, Katowice, Poland}
\author{M. Kirsanov}
\affiliation{INR RAS, Moscow, Russia}
\author{J. Kisiel}
\affiliation{Institute of Physics, University of Silesia, Katowice, Poland}
\author{I. Kochanek}
\affiliation{INFN - Laboratori Nazionali del Gran Sasso, Assergi, Italy}
\author{J. Lagoda}
\affiliation{National Centre for Nuclear Research, Otwock/Swierk, Poland}
\author{A. Menegolli}
\affiliation{Dipartimento di Fisica Universit\`{a} di Pavia and INFN, Pavia, Italy}
\author{G. Meng}
\affiliation{Dipartimento di Fisica e Astronomia Universit\`{a} di Padova and INFN, Padova, Italy}
\author{C. Montanari}
\affiliation{Fermi National Accelerator Laboratory, Batavia, IL 60510, USA}
\affiliation{On leave of absence from INFN Pavia, Pavia, Italy}
\author{S. Otwinowski}
\affiliation{Department of Physics and Astronomy, UCLA, Los Angeles, USA}
\author{C. Petta}
\affiliation{Dipartimento di Fisica e Astronomia Universit\`{a} di Catania and INFN, Catania, Italy}
\author{F. Pietropaolo}
\affiliation{CERN, European  Organization for Nuclear Research, 1211 Geneva 23, Switzerland}
\affiliation{On leave of absence from INFN Padova, Padova, Italy}
\author{A. Rappoldi}
\affiliation{Dipartimento di Fisica Universit\`{a} di Pavia and INFN, Pavia, Italy}
\author{G.L. Raselli}
\affiliation{Dipartimento di Fisica Universit\`{a} di Pavia and INFN, Pavia, Italy}
\author{M. Rossella}
\affiliation{Dipartimento di Fisica Universit\`{a} di Pavia and INFN, Pavia, Italy}
\author{C. Rubbia}
\affiliation{INFN - Laboratori Nazionali del Gran Sasso, Assergi, Italy}
\affiliation{CERN, European  Organization for Nuclear Research, 1211 Geneva 23, Switzerland}
\affiliation{GSSI, L’Aquila, Italy}
\author{P. Sala}
\affiliation{Present address: Fermi National Accelerator Laboratory, Batavia, IL 60510, USA}
\author{A. Scaramelli}
\affiliation{Dipartimento di Fisica Universit\`{a} di Pavia and INFN, Pavia, Italy}
\author{F. Sergiampietri}
\affiliation{CERN, European  Organization for Nuclear Research, 1211 Geneva 23, Switzerland}
\affiliation{Present  address: IPSI-INAF Torino, Torino, Italy}
\author{D. Stefan}
\affiliation{Dipartimento di Fisica Universit\`{a} di Milano and INFN, Milano, Italy}
\author{M. Szarska}
\affiliation{H. Niewodniczanski Institute of Nuclear Physics, Polish Academy of Science, Krakow, Poland}
\author{M. Terrani}
\affiliation{Dipartimento di Fisica Universit\`{a} di Milano and INFN, Milano, Italy}
\author{M. Torti}
\affiliation{Dipartimento di Fisica Universit\`{a} di Milano Bicocca and INFN, Milano, Italy}
\author{F. Tortorici}
\affiliation{Dipartimento di Fisica e Astronomia Universit\`{a} di Catania and INFN, Catania, Italy}
\author{F. Varanini}
\affiliation{Dipartimento di Fisica e Astronomia Universit\`{a} di Padova and INFN, Padova, Italy}
\author{S. Ventura}
\affiliation{Dipartimento di Fisica e Astronomia Universit\`{a} di Padova and INFN, Padova, Italy}
\author{C. Vignoli}
\affiliation{INFN - Laboratori Nazionali del Gran Sasso, Assergi, Italy}
\author{H. Wang}
\affiliation{Department of Physics and Astronomy, UCLA, Los Angeles, USA}
\author{X. Yang}
\affiliation{Department of Physics and Astronomy, UCLA, Los Angeles, USA}
\author{A. Zalewska}
\affiliation{H. Niewodniczanski Institute of Nuclear Physics, Polish Academy of Science, Krakow, Poland}
\author{A. Zani}
\affiliation{Dipartimento di Fisica Universit\`{a} di Milano and INFN, Milano, Italy}
\author{K. Zaremba}
\affiliation{Institute of Radio Electronics, Warsaw University of Technology, Warsaw, Poland}
\collaboration{ICARUS Collaboration at Gran Sasso}
\date{\today}

\begin{abstract}

We present the result of a search for inelastic boosted dark matter using the data corresponding to an exposure of 0.13 kton$\cdot$year, collected by the ICARUS T-600 detector during its 2012--2013 operational period at the INFN Gran Sasso Underground National Laboratory. The benchmark boosted dark matter model features a multi-particle dark sector with a U(1)$'$ gauge boson, the dark photon. The kinetic mixing of the dark photon with the Standard Model photon allows for a portal between the dark sector and the visible sector. 
The inelastic boosted dark matter interaction occurs when a dark matter particle inelastically scatters with an electron in the ICARUS detector, producing an outgoing, heavier dark sector state which subsequently decays back down to the dark matter particle, emitting a dark photon. The dark photon subsequently couples to a Standard Model photon through kinetic mixing. The Standard Model photon then converts to an electron-positron pair in the detector.  This interaction process provides a distinct experimental signature that consists of a recoil electron from the primary interaction and an associated electron-positron pair from the secondary vertex. After analyzing 4,134 triggered events, the search results in zero observed events.
Exclusion limits are set in the dark photon mass and coupling ($m_X, \epsilon$) parameter space for several selected optimal boosted dark matter mass sets and cover previously unexplored parameter space. 

\end{abstract}

\maketitle
\let\clearpage\relax
\section{Introduction} \label{introduction}
The dark matter (DM) hypothesis is greatly motivated by the gravitationally measured mass to the visible mass ratio found by cosmological observations at various scales. The observations include the rotational dynamics of galaxies~\cite{GalaxyRotationCurves}, the dynamics of galaxy clusters~\cite{Zwickypaper,BulletClusterpaper}, and the overall matter distribution of large-scale structure formation seen in the Cosmic Microwave Background (CMB) power spectrum~\cite{PlanckResults2018,WMAP_2013,Salucci_2019}. The general properties of a DM particle model can be summarized under the $\Lambda$CDM model~\cite{lambdaCDMCondon_2018,Profumo2017}: 1) DM has to be dark, having small to no electromagnetic charge, 2) DM's interactions are gravitationally dominated, and 3) DM's bulk velocity has to be non-relativistic. 

The Weakly Interacting Massive Particle (WIMP) paradigm (see, e.g., Refs.~\cite{Steigman:1984ac,Jungman:1995df,Feng:2022rxt}) is a benchmark theory for single-constituent DM and direct DM detection experiments~\cite{WIMPcrossmass_paper}. As of today, no conclusive WIMP signal has been detected via its hypothetical non-gravitational interactions, excluding a large area of parameter space where the WIMP mass is larger than $O$(GeV) (see, e.g., Fig. 3 in Ref.~\cite{WIMPcrossmass_paper} and references therein). When these WIMP search efforts are extended to lower mass scales, DM masses at the sub-GeV level require detector sensitivity for DM-nucleon interactions around sub-keV or below, challenging currently available detector technologies to measure such low-energy depositions in the detector medium.

An alternative approach to address these challenges is to ``indirectly'' search for halo (i.e., dominant) DM via subdominant DM components within multi-component DM scenarios~\cite{DirectDarkMatter}. In such scenarios, the halo DM component (possibly WIMP scale) has suppressed couplings to Standard Model (SM) particles but can pair-annihilate into typically lighter, subdominant DM particles with a significant boost factor. These scenarios further allow a subdominant DM species to interact with SM particles through a new U(1)$'$ gauge mediator, such as a dark photon $X$~\cite{Okun:1982xi,Galison:1983pa,Holdom:1985ag}, which couples the dark sector to the SM via kinetic mixing with the SM photon.

In this context, this paper presents a search for inelastic boosted dark matter (iBDM)~\cite{Kim:2016zjx,Giudice:2017zke} via its interaction with electrons, producing a heavier, excited dark sector state which results in an electron-positron pair final states, displaced from the primary vertex, using the ICARUS T-600 detector at the INFN Gran Sasso Underground National Laboratory (LNGS)~\cite{ICARUSinitial}. An earlier iBDM search effort was made at COSINE-100~\cite{COSINE-100:2018ged}, a ton-scale DM detector. In contrast, the ICARUS-T600 detector is a multi-hundred-ton liquid argon time projection chamber (LArTPC) that operated successfully at LNGS to study the beam-produced neutrinos sent by CERN SPS, CERN Neutrinos at Gran Sasso program (CNGS)~\cite{ICARUSpaper}, and the atmospheric neutrinos~\cite{universe501001}. 
A low energy threshold of 200 MeV and excellent energy reconstruction capabilities through the $dE/dx$ analysis of particle tracks in the detector~\cite{ICARUSTrackReconstructionPaper} enable an accurate particle identification, which is essential for distinguishing iBDM signal from backgrounds.

This paper is organized as follows.  An overview of the BDM model and the iBDM process are discussed in Sec.~\ref{BDM Theory}. The ICARUS T-600 detector and the data used for this study are presented in Sec.~\ref{The ICARUS Detector}. The iBDM signal simulation at the theoretical level and the detailed detector level are described in Sec.~\ref{Simulation}.
The detailed analysis procedure and the methodology, including that for determining the optimal parameter sets, are presented in Sec.~\ref{Analysis}.
Section~\ref{Backgrounds} describes the background reduction and estimate. 
Finally, Sec.~\ref{Results} presents the final results of the search, with the conclusions following in 
Sec.VIII.
\section{Boosted Dark Matter} \label{BDM Theory}
Phenomenology arising from relativistic dark matter produced in the present universe differs significantly from the predictions of conventional WIMP-type dark matter models. Over the past decade, numerous mechanisms for boosting non-relativistic dark matter have been proposed, and extensive studies have explored the phenomenology of non-minimal dark sector scenarios (see, e.g., Ref.~\cite{Berger:2022cab} and references therein). In this work, we use the concept of inelastic dark matter within two-component dark matter scenarios as a benchmark for illustration.

Models of two-component BDM~\cite{indirectDMdetection_2014,Kim:2016zjx,Giudice:2017zke} operates with a multi-particle dark sector, incorporating a non-relativistic, WIMP-like component $\chi_0$ (mass $m_0$), but also a lower mass component $\chi_1$ (mass $m_1$). Their relic abundances are described by the assisted freeze-out mechanism~\cite{AssistedFreezeout_2012}, demonstrating that the BDM models can assimilate dynamics required by DM in the early universe to create the large-scale structure seen in the CMB. The production mechanism for boosted $\chi_1$ involves $\chi_0$ self-annihilation $\bar{\chi}_0 \chi_0\rightarrow \bar{\chi}_1 \chi_1$ in the present universe, allowing $\chi_1$ to acquire energy equal to the mass of $\chi_0$ (i.e., significant boost factor). Taking the dominant production region as the galactic center and assuming the Navarro-Frenk-White DM halo profile~\cite{Navarro:1995iw}, the $\chi_1$ flux becomes~\cite{indirectDMdetection_2014}

\begin{equation}
    \label{eqn:chiflux}
    F_{\chi_1}=1.6 \times 10^{-4}{\rm cm}^{-2}{\rm s}^{-1}\left (\frac{\langle \sigma v \rangle_{0\rightarrow 1}}{5 \times 10^{-26}{\rm cm}^3{\rm s}^{-1}}\right ) \left ( \frac{\rm GeV}{m_0}\right )^2
\end{equation}
where $\langle \sigma v \rangle_{0 \rightarrow 1}$ is the thermally-averaged cross-section of the self-annihilation process $\bar{\chi}_0 \chi_0 \rightarrow \bar{\chi}_1 \chi_1$. 

In addition to $\chi_1$, models of iBDM further hypothesize that there is a heavier unstable state, $\chi_2$ (mass $m_2$). Hence, the iBDM Lagrangian includes the following relevant operators~\cite{Kim:2016zjx,Giudice:2017zke}

\begin{eqnarray}
    -\mathcal{L} &\supset& \epsilon e \bar{f}\gamma^\mu f X_\mu + g_{11}\bar\chi_1\gamma^\mu\chi_1 X_\mu \nonumber \\
    &&+ g_{12}\bar\chi_2\gamma^\mu\chi_1 X_\mu + h.c.,  \label{eqn:lagrangian} 
\end{eqnarray}
where $f$ is the SM fermion field including electron,
$X_\mu$ is the dark photon of mass $m_X$, $g_{11}$, and $g_{12}$ are the flavor-conserving and flavor-changing couplings, respectively, and $\chi_{1(2)}$ are the dark sector spinors (fermionic DM scenario).  
Under this scenario, $\chi_2$ can be produced by energetic or boosted $\chi_1$ inelastically scattering with the detector medium. The dark sector mass hierarchy is assumed to be $m_0\gg m_2>m_1$, where each subscript denotes the corresponding dark sector particles, including BDM. Once produced in the detector, $\chi_2$ decays back to a $\chi_1$ and a dark photon which subsequently decays to SM leptonic pairs via kinetic mixing between the dark photon and the SM photon~\cite{TheDarkPhoton}. 

\begin{figure}
    \centering 
    \includegraphics[width=1.00\linewidth]{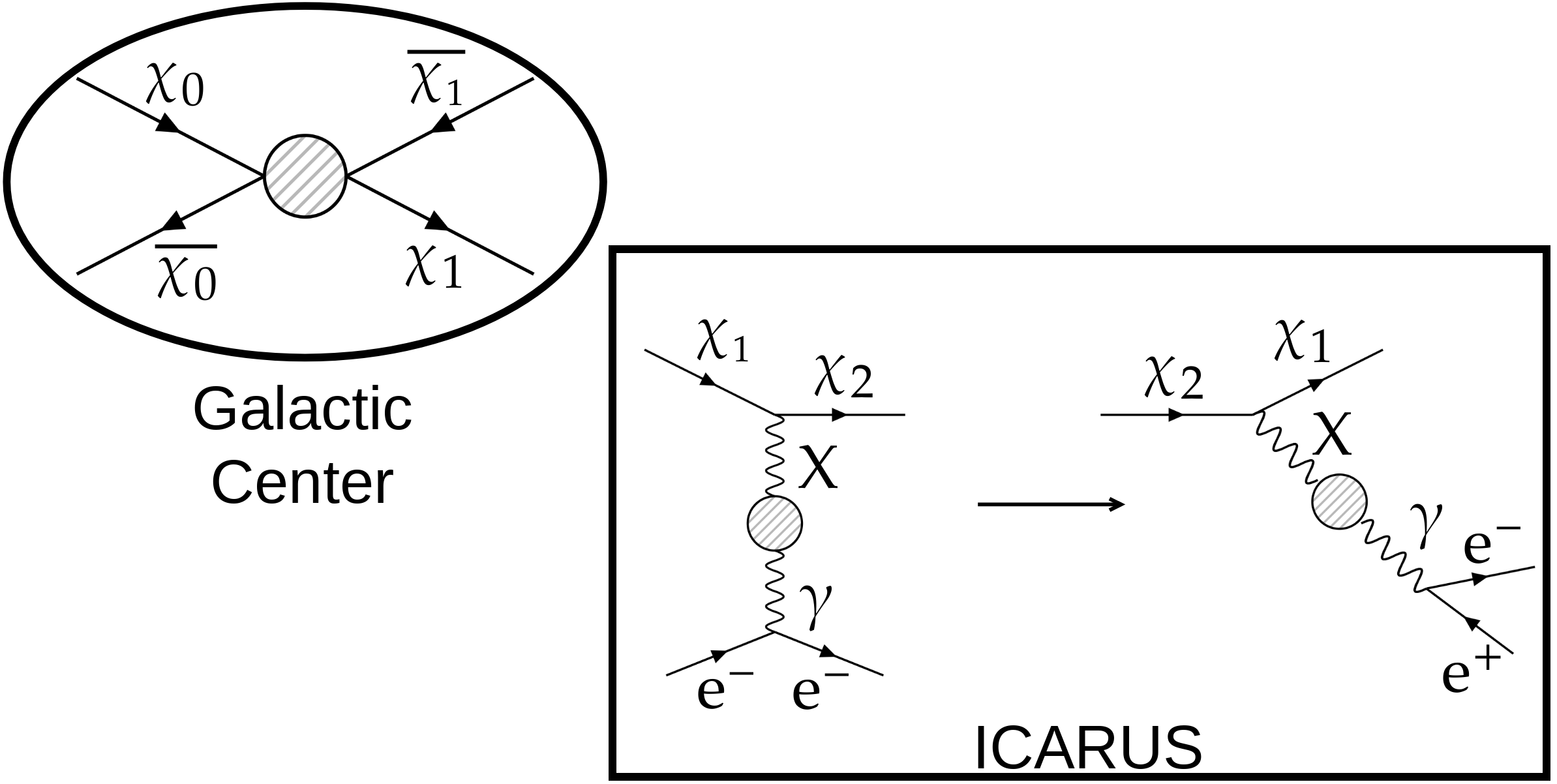}
    \hfill
    \caption{Pictorial representation of the production of $\chi_1$ at the galactic center, leading to an iBDM interaction in the ICARUS detector. The primary interaction can theoretically involve an electron (this work) or proton of the LAr atom.}
    \label{fig:iBDMfigure}
\end{figure}

For this paper, we study the primary interaction of boosted $\chi_1$ with an electron 
\begin{equation}
    \chi_1 e^-\rightarrow \chi_2 e^-_R,
\end{equation}
where $e^-_R$ is the recoil electron, and the subsequent secondary interaction is the decay of the $\chi_2$ to electron-positron pairs 
\begin{equation}
\chi_2 \to \chi_1(X^{(*)}) \to \chi_1 e^-e^+.
\end{equation}
Here the $X^{(*)}$ symbol includes the possibility of the $\chi_2$ decay via an on-(off-)shell dark photon, depending on the underlying mass hierarchy among particles involved in the decay process. In the off-shell case, $\chi_2$ often decays at a location displaced from the primary scattering point~\cite{Kim:2016zjx}. The concept of searching for iBDM events at LArTPC detectors in this manner has been explored in phenomenological literature~\cite{Kim:2016zjx,Chatterjee:2018mej,Kim:2020ipj,De_Roeck_2020}.
Figure~\ref{fig:iBDMfigure} shows the process of the creation of $\chi_1$ at the galactic center leading to an iBDM interaction in the ICARUS detector, symbolically with Feynman diagrams. The visible particles in the detector are the primary interaction recoil electron ($e^-_R$) and the secondary interaction electron-positron pair ($e^-e^+$). 

\section{The ICARUS Experiment} \label{The ICARUS Detector}
This section describes the ICARUS experiment at the LNGS underground laboratory, the relevant detector performance parameters, and the data sample used for the iBDM search analysis.
\subsection{Detector overview}
The Liquid Argon Time Projection Chamber technology was first proposed in 1977 by Carlo Rubbia~\cite{Rubbia:1977zz}, as an alternative to Cherenkov detectors for astroparticle physics experiments, proton decay searches, and neutrino physics, providing both precision 3D tracking and calorimetric energy measurement capabilities.

Charged particles that traverse the pure LAr medium ionize the Ar atoms and simultaneously produce scintillation light from the decay of the excited Ar atoms and the ionization charge recombination. 
The ionization electrons drift under a uniform electric field to a set of anode wire planes, where they are collected within the set readout time window. Thanks to the low transverse diffusion of charges in LAr, the images of the ionizing tracks are preserved along the drift direction, allowing the precise tracking of the charged particle trajectory. The wire plane signal coupled with the corresponding prompt scintillation light signal collected by the photon detection system allows for full 3D track reconstruction of the recorded events inside the readout time window, initiated by a scintillation light trigger system. The position of the tracks along the drift direction is determined by the absolute time of arrival after the trigger, combined with the electron drift velocity, $v_D \sim 1.6$~mm/$\mu$s under the uniform $\sim 500$ V/cm electric field.

The ICARUS T-600 LArTPC detector, which contains a total of 760 tons of ultra-pure LAr, operated and took data at the Gran Sasso Underground National Laboratory of INFN~\cite{ICARUSbible,ICARUSinitial} in 2011 -- 2013. The detector consists of two independent T300 cryostat modules with the internal dimensions {\rm $3.6~{\rm m\,(W)} \times 3.9~{\rm m\,(H)}\times 19.9~{\rm m\,(L)}$}.
Each module includes two TPCs which share a common semi-transparent cathode plane in the center of the module. The TPC anode is composed of three parallel wire planes, 3~mm apart, positioned on either side of the cathode plane and facing inward to the active volume with the 1.5~m drift path. A total of 53,248 wires with lengths up to 9~m are installed in the ICARUS-T600 detector, providing precision tracking capability. 

With an appropriately graded bias voltage, the first two planes, Induction-1 and Induction-2 planes, provide signals in a non-destructive manner. The charge of the ionization electrons is finally collected by the last plane, the Collection plane, for the measurement of the particle energy deposition. The wires of the Induction-1 plane face the cathode and run horizontally along the length of the detector. The Induction-2 wire plane, 3~mm behind the Induction-1, has a wire orientation of $60^{\circ}$ with respect to the orientation of the Induction-1 wires. The wires of the Collection plane are oriented $-60^{\circ}$ with respect to the Induction-1 wire direction.
Combining the arrival time of the signal on the wire planes with the corresponding wire position, the adopted stereo-angle wire configuration provides three bi-dimensional projections of any charged particle tracks.
The 3~mm wire pitch and the 3~mm wire plane separations provide the 3D event reconstruction with $\sim {\rm 1~mm}^3$ spatial resolution~\cite{ICARUSbible,ICARUSTrackReconstructionPaper}. 

The TPC active volume between the cathode and anode is enveloped by the field cage which is composed of 29 equally spaced racetrack-shaped stainless steel electrodes~\cite{ICARUSbible}. The distance between two neighboring electrodes is $49.6$~mm. Each electrode is biased to create a uniform drift electric field of 500 V/cm along the $\sim1.5$~m drift distance from the cathode to the anode. Globally, the total active LAr mass amounts to 476 tons for the entire ICARUS T-600 detector.

The scintillation light is collected by the $8"$ diameter Photo Multiplier Tubes (PMTs), coated with a wavelength shifter to allow for the detection of vacuum ultra-violet (VUV) scintillation light that has the wavelength of $\lambda$~=~128~nm. Two arrays of 20 and 54 PMTs are installed in the $1^{\rm st}$ and $2^{\rm nd}$ T300 module, respectively, behind the anode wire planes and outside of the active volume, to provide the initial time (i.e., $t_0$) of an event and the trigger signal.
The readout electronics was designed to allow continuous readout, digitization, and independent waveform recording of the signal from each wire of the TPCs and the PMTs.

LAr was continuously filtered and recirculated to remove the electronegative impurities, mainly oxygen, which capture the ionization electrons during the drift.
The residual LAr impurities were kept below 50~ppt O$_2$ equivalent throughout the entire data-taking period, corresponding to $\sim 12\%$ maximum charge attenuation at the longest drift distance~\cite{LArpurity}. The LAr purity was monitored by measuring the attenuation of the through-going cosmic muon tracks along the drift direction to correct the charge signal on the TPC wires.  
The fine granularity of the detector and the resulting high resolution allow for a precise reconstruction of the event topology and the recognition of the particles produced in an interaction in LAr. The event reconstruction is completed by calorimetric measurements via $dE/dx$ ionization signal over a very wide energy range, from MeV to several tens of GeV.
The particle identification process is performed by studying the event topology and the local energy deposition per unit length, $dE/dx$. Muons and electrons exhibit the minimum ionizing particle (m.i.p.) characteristic [($dE/dx$)$_{\rm m.i.p.}\sim$~2~MeV/cm].

Electrons are fully identified by the characteristic electromagnetic (e.m.) showering. They are well distinguished from $\pi^0$s via the $\gamma$ reconstruction, $dE/dx$ signal comparison, and $\pi^0$ invariant mass measurement. This feature guarantees a powerful identification of the electron neutrino charged-current (CC) interactions while rejecting the neutral current (NC) interactions to a negligible level.  

The ICARUS T600 trigger system used for the collection of cosmic events out of the CNGS beam spill, is based on the scintillation light signal collected by the PMTs located behind the wire planes. 
The analog sum of the signals from the PMTs in the same chamber was used, with a defined photo-electron (phe) discrimination threshold for each TPC chamber, $\sim100$ phe in the west cryostat and $\sim200$ phe in the east cryostat to account for the different number of the deployed PMTs, and the trigger was provided by the coincidence of the PMT sum signals of the two adjacent chambers in the same module.
The efficiency of the PMT sum signal depends on the total energy deposit in the event and the distance from the event to the cryostat walls on which the PMTs are mounted. 
The efficiency is minimally affected by the smaller number of PMTs in the first module.
Overall, the PMT sum trigger efficiency varies between 80\% -- 100\% for the events that deposit energy in the detector greater than 200 MeV ($E_{\rm dep} > 200 $ MeV)~\cite{ICARUSTrigger}.

\subsection{The Data Sample}
The data set used for the iBDM search presented in this paper corresponds to 0.13~kton$\cdot$year exposure. This is part of the data collected by ICARUS in the 2012 -- 2013 operation, a total exposure of 0.43~kton$\cdot$year. 
The detector was situated under the Gran Sasso mountain, covered by $\sim$\,3,400 meter water equivalent (m.w.e) rock, greatly suppressing cosmic rays and allowing for a highly sensitive study of neutrino interactions~\cite{universe501001}. 
It is important to note that due to the cosmic flux suppression, each triggered event corresponds to a single interaction in the entire ICARUS T600 detector, with negligible contamination from uncorrelated tracks crossing the detector within the drift time.

ICARUS neutrino events are categorized as either the neutrinos delivered via the CNGS beam or the atmospheric neutrinos, for both $\nu_e$ and $\nu_{\mu}$ CC interaction studies. The analysis of both the CNGS beam and atmospheric neutrino events demonstrated the unique capability of triggering, collecting, and accurately reconstructing neutrino events using the ICARUS LArTPC detector. 

In order to collect atmospheric neutrino events, a cosmic trigger was implemented outside the CNGS beam spill window. To mitigate the high-energy cosmic muon background, a software filter was developed to identify the $\nu_e$CC and $\nu_{\mu}$CC interactions~\cite{universe501001}. The filter algorithm uses the charge signals on the Collection wires, i.e. the hits, and groups the hits into clusters based on their relative distances. The cluster with the greatest number of hits in the event is then identified by the thresholds imposed on the spatial and calorimetric properties~\cite{universe501001}. Large clusters of hits indicate the presence of an e.m. shower from an electron track in the event. The $\nu_e$CC nuclear interaction ($\nu_e n\rightarrow pe^-$) produces an outgoing electron that can shower, making the filtering procedure  appropriate for the identification of $\nu_e$CC interactions.

The adopted software filter identifies $\nu_e$CC events with an efficiency just above 80\% (see Table~\ref{table:tableneutrinostudy}). Due to the filter's bias for the e.m. shower recognition, however, its efficiency on $\nu_\mu$CC is significantly lower. In addition, the filter was designed to reject straight tracks, which have a high probability of being cosmic muons. Therefore, the muons from $\nu_{\mu}$CC interactions could also be filtered out, further reducing the efficiency.

\begin{table}[b]  
  \centering
  \begin{tabular}{llll}
    \toprule
    Stages of the Analysis & $\nu_{\mu}$CC & $\nu_e$CC  \\ \midrule
    $N_{\rm evt}^{\rm Expected}$ per kton$\cdot$year & 96.2 & 78.2\\ 
    $N_{\rm evt}^{\rm Expected}$ for 0.43~kton$\cdot$year exposure& 41.4 & 33.7\\
    Including the Fiducial Volume & 37.8 & 30.8\\
    Including the Deposited Energy $>$ 200 MeV & 24.9 &  24.2\\
    Filter Efficiency($\xi_{\rm filter}$) & 25.7\% & 81.4\% \\
    Including the filter efficiency & 6.4 & 19.7 \\ 
    Trigger efficiency($\xi_{\rm trigger}$) & 86.7\% & 84.7\% \\ 
    Including the trigger efficiency & 5.5 & 16.7 \\ 
    Including scanning efficiency($\xi_{\rm scanning}$= 80\%) & 4.4 & 13.3\\ \hline
    {\bf Final $N_{\rm evt}^{\rm Expected}$} & {\bf 4.4} & {\bf 13.3} \\ \hline\hline
    {\bf Number of observed events} & {\bf 6} & {\bf 8} \\ \bottomrule
  \end{tabular}
  \caption{Comparisons of the expected number of events, $N_{\rm evt}^{\rm Expected}$ at each stage of the atmospheric neutrino study from Ref.~\cite{universe501001}. Each row successively applies the detector acceptance and the selection efficiencies, with the final expected number of neutrino events at the application of the scanning efficiency highlighted. The actual number of observed events in the bottom row, is consistent with the final expected number of events within the statistical uncertainty.}
  \label{table:tableneutrinostudy}
\end{table}

Each row in Table~\ref{table:tableneutrinostudy} refers to a step in the atmospheric neutrino event selection. Through Monte Carlo simulations, the expected numbers of $\nu_e$ and $\nu_\mu$ interactions per kton$\cdot$year of exposure are estimated. Every experimental cut and the corresponding efficiency are then applied successively to obtain the final number of expected events per kton$\cdot$year. 

The final step of the atmospheric neutrino study was the identification of the neutrino interactions in the data sample that passed the filter software procedure.  The identification was performed visually by scanners who were trained using the simulated $\nu_e$ and $\nu_\mu$ events through the full, detailed detector simulation, for topological study in the event display.  All three wire plane views of the detailed simulated neutrino events are studied for topology and calorimetric visual recognition. All the atmospheric neutrino candidates with a clear interaction vertex within the fiducial volume, i.e. at a distance larger than 5~cm inward from the extreme edges of the active volume, have been considered. The $\nu_e$CC events were identified by the presence of a clear e.m. shower from the primary vertex, with a $dE/dx$ signal at the beginning of the shower, evaluated in the first few wires, compatible with a m.i.p. The $\nu_\mu$CC events were selected requiring a long track (at least 1 m) from the primary vertex. At the end of the scanning, a combined 17.7 neutrino events were estimated to be identified, and a total of 14 events were observed in the data~\cite{universe501001}.

The signature of an iBDM interaction in the detector is two distinct but associated e.m. showers, one from the recoil electron in the primary interaction and the other from the electron-positron pair in the secondary interaction.
Since the data set filtered for the atmospheric neutrino analysis requires the presence of at least one e.m. shower, iBDM signal events sought in this analysis must be contained in this data. 
The same criteria used for the identification of atmospheric neutrinos are applied to the identification of iBDM candidates. The parameters in Table~\ref{table:tableneutrinostudy} that must be addressed specifically for iBDM candidates, however, are the expected number of iBDM events for the dataset exposure, the efficiency of the filter to the iBDM signal, and the scanning efficiency for the iBDM signal events. Each of these parameters is addressed with the analysis of the detailed Monte Carlo simulated iBDM signal sample.
These are described in detail in Sec.~\ref{Analysis}.
\section{iBDM Signal Simulation} \label{Simulation}
The iBDM simulation consists of two components: the Monte Carlo event generator and the detailed full detector iBDM signal simulation. The iBDM signal simulation studies are performed to (1) establish the selection criteria, (2) evaluate the event selection criteria efficiency ($\xi_{\rm criteria}$), and (3) train scanners on the different iBDM signal topologies present in the optimal parameter sets that satisfy the selection criteria. 

The iBDM event generator takes as the input the values for the seven free parameters of the BDM model~\cite{indirectDMdetection_2014} -- the three dark sector particle masses including DM, $m_0$, $m_1$, and $m_2$, the dark photon kinetic mixing $\epsilon$ and mass $m_X$, and the interaction couplings $g_{11}$ and $g_{12}$ in generating iBDM events. 

The output of the simulation is the kinematic truth information of the simulated iBDM events. The iBDM event generator was developed in-house and was also used for DUNE BDM detector sensitivity studies~\cite{De_Roeck_2020}. Through the application of selection criteria, the kinematic information is used to obtain the optimal mass parameter sets that maximize the number of expected events in the dark photon parameter space. 

The kinematic truth information of the Monte Carlo-generated events that use the optimal mass parameter sets becomes the input to the detailed full detector simulation. A GEANT4-based~\cite{Geant4} detector simulation uses the detailed description of the detector geometry and the medium LAr and simulates the amount of the ionization charge produced by an iBDM interaction in the detector. 
The charge information from the GEANT4 simulation becomes the input for the wire simulation used for ICARUS at Gran Sasso. This wire simulation incorporates realistic detector efficiencies, such as the trigger and filter efficiencies. 

\begin{figure}[b]
    \centering
    \includegraphics[width=1.0\linewidth,height=0.6\linewidth]{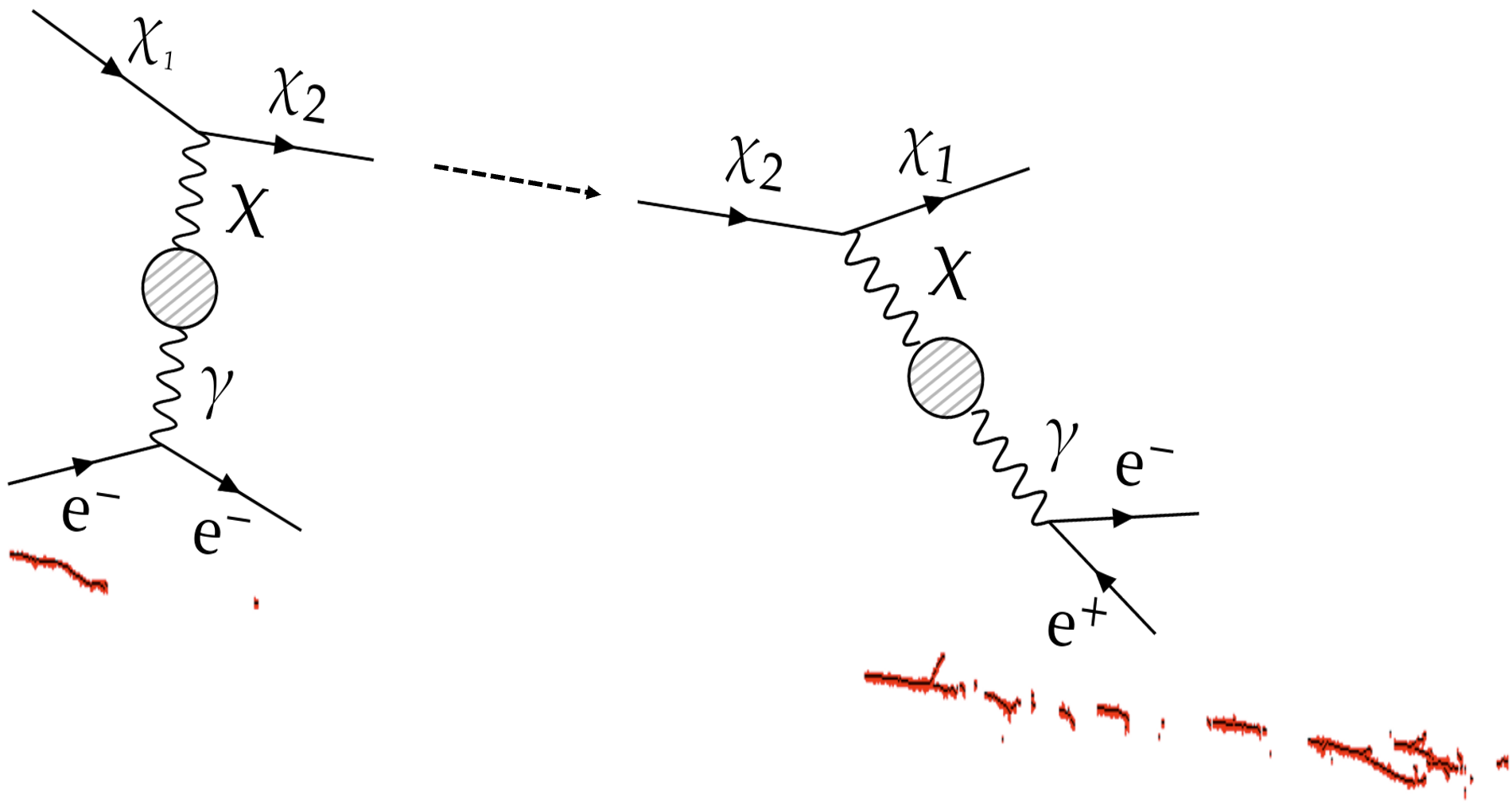}
    \caption{An event display of a simulated iBDM event. Red dots indicate the hits registered in the detector with the primary and secondary interactions indicated in the corresponding Feynman diagrams above the hit. The dashed arrow indicates that $\chi_{2}$ state moves a certain distance in the detector before emitting the dark photon $X$ and decaying back to $\chi_{1}$.  In the detector, the showering of the $e^{+}e^{-}$ pair could continue to propagate through the detector.}
    \label{fig:iBDMeventdisplay}
\end{figure}

The data used for this analysis is the data selected through the filter algorithm for the atmospheric neutrino study. Since the filter algorithm is already applied to the data, the filter efficiency is evaluated for each of the iBDM parameter sets in this study, using the detailed simulation.  The wire information is available at the end of the iBDM signal simulation, allowing for the study of iBDM track topologies and energy deposits, as well as the training in visual event scanning. Figure~\ref{fig:iBDMeventdisplay} shows an example iBDM signal event.  
The red points indicate the wire ``hits'', and the Feynman diagrams depict which wire hits correspond to the primary and secondary interactions.

\section{Analysis} \label{Analysis}
This section presents the iBDM simulation analysis to establish event selection criteria (Sec.~\ref{selectioncriteria}), identify optimal ($m_0,m_1,m_2$) mass parameters that maximize the coverage in the dark photon parameter space (Sec.~\ref{simulationstudyI}), and evaluate the detector performance in the optimal parameter space with the wire signals from the full detector simulation (Sec.~\ref{simulationstudyII}). Finally, the real data analysis methodology is presented in Sec.~\ref{realdataanalysis}, including the uncertainties associated with the measurements that impose the selection criteria in Sec.~\ref{EvaluationofUncertainties}). 
\subsection{Selection Criteria}\label{selectioncriteria}
The iBDM signal topology for this analysis is unique thanks to the presence of an $e_R^-$ initiated e.m. shower followed by an associated $e^+e^-$ pair e.m. shower produced in the associated secondary interaction. The full containment of the primary and secondary iBDM interaction is essential for the complete identification of an iBDM interaction.

Electrons, positrons, and muons have the characteristic m.i.p. $dE/dx$ signature. Since electrons and positrons produce showers after they travel some distance through the LAr, the m.i.p. signature is present at the beginning of the travel for tracks produced by these particles. Therefore, the primary interaction electron and the secondary interaction electron-positron pair must be displaced from each other at a minimum distance of $\sim3$ cm in order to identify accurately identify the m.i.p. signature of the primary interaction track and the two-m.i.p. signature of the secondary interaction track.

The unique capability of the detector in distinguishing electron and photon imitated e.m. showers~\cite{ICARUSpaperLSND,ICARUSpaper}, together with the additional requirement that no muon or no hadronic activities are present in the event, enable reducing backgrounds from all sources to a negligible level, with a minimal impact on the signal search efficiency as is described in Sec.~\ref{Backgrounds}. 

Accounting for the measured PMT trigger efficiency $\> 80\%$~\cite{ICARUSTrigger} in the events with energy deposition greater than 200~MeV in the detector active volume, a total energy deposition threshold for the iBDM $E_{\rm thres}=200$ MeV has been set to keep the same trigger efficiency $\xi_{\rm trigger}\geq0.8$.

In summary, events that satisfy the following selection criteria in succession are identified as iBDM candidate events:
\begin{enumerate}
    \item The primary and secondary interaction vertex is contained within the fiducial volume, defined as 5~cm inward from the extreme edge of the detector active volume.
    \item The primary and secondary interaction vertices are at least 3~cm apart. This helps in the identification of the recoil electron from the primary interaction as a minimum m.i.p. near the vertex.
    \item The total visible energy $E_{\rm tot}$ by the recoil electron and by the $e^+e^-$ pair, i.e., $E_{\rm tot}= E_R + E_{e^-e^+}$, is above the 200~MeV threshold: $E_{\rm tot}\geq E_{\rm thres}=200$~MeV.
    \item No hadronic activity, no muons, and no charged particle entering into the active volume from outside are present in the event.
\end{enumerate}
The selection criteria 1 -- 3 are imposed on simulated iBDM events in Sec.~\ref{simulationstudyI} and Sec.~\ref{simulationstudyII} for the evaluation of the detector sensitivity in both the mass parameter space ($m_0,m_1,m_2$) and dark photon parameter space ($m_X,\epsilon$). Criterion 4 is imposed in the real data scanning to discriminate a neutrino CC interaction from an iBDM interaction and to identify the cosmic ray or atmospheric neutrino interactions outside the detector entering the fiducial volume and potentially mimicking the iBDM signal.

Previous search results (e.g., Refs.~\cite{OHare:2021utq,WIMPcrossmass_paper}) based on the standard WIMP scenarios impose minimal constraints on $m_0$ for the dominant halo dark matter $\chi_0$~\cite{Agashe:2014yua}. $\chi_1$ and $\chi_2$ are subdominant and unstable dark-sector species, respectively, leaving $m_1$ and $m_2$ to remain largely unconstrained.
By imposing the selection criteria 1 -- 3 above on simulated iBDM events and imposing a 90\% C.L. limit at the edge of the dark photon ($m_X,\epsilon$) space, a range of ($m_0,m_1,m_2$) that maximize the number of expected iBDM signal events under these conditions are identified as described in Sec.~\ref{simulationstudyI}, assuming the 100\% event selection efficiency. Sec.~\ref{simulationstudyII} presents the global detection efficiency for the optimal mass set determined in Sec.~\ref{simulationstudyI}.

\subsection{Optimal Parameter Set Determination Methodology}
Each BDM parameter is constrained by current cosmological observations and Earth-based experiments such as beam dumps (e.g., nu-Cal~\cite{nucal2011}, E141~\cite{E141}), collider/fixed target (e.g., NA64(e)~\cite{NA64e}, NA48/2~\cite{NA48}) and other types of experiments such as ($g-$2)$_e$~\cite{gminus2}. The parameters can be grouped into the mass parameters $(m_0,m_1,m_2)$, the interaction coupling parameters ($g_{11}, g_{12}$), and the dark photon parameters ($m_X, \epsilon$).
In our analysis, we consider two chiral fermion scenarios as benchmark physics cases~\cite{Giudice:2017zke}, while our search results remain applicable to generic iBDM models.
In this type of scenario, the relative proportion of the interaction coupling parameters depends on the eigenvalues of the DM masses ($m_1, m_2$) which consist of the Majorana mass component and the Dirac mass component related to $g_{11}$ and $g_{12}$, respectively. 
The sum of the two interaction coupling parameters squared normalized by the dark-sector cooling $g_D$ is one~\cite{Giudice:2017zke}:
\begin{equation}
\left(\frac{g_{11}}{g_D}\right)^2+\left(\frac{g_{12}}{g_D}\right)^2=1. 
\end{equation}
With $g_D=1$ as a benchmark choice, in the scenario where the Dirac mass dominates, $g_{11}$ is suppressed, namely $g_{11}\sim 0$, while the iBDM interaction dominates $g_{12}\sim 1$. 
This leaves the DM mass parameters and the dark photon parameters free. 

While there are many ways of exploring the iBDM model space due to the multi-parameter nature of iBDM, we decided to focus on the maximum reachable space of both the dark sector mass parameters ($m_0,m_1,m_2$) and the dark photon parameters ($m_X,\epsilon$) for the ICARUS experiment. The procedure for determining these spaces follows the steps below, successively:
\begin{enumerate}
    \item Fix the dark photon parameters set $(m_X,\epsilon)$ at the present exclusion limit 
    \[(m_X,\epsilon)_{\rm limit}=(12~{\rm MeV},~0.0008)\] 
    and identify the optimal $(m_0,m_1,m_2)$ mass sets that maximize the number of expected events at the given ($m_X,\epsilon$)$_{\rm limit}$, passing the selection criteria 1 -- 3 presented in Sec~\ref{selectioncriteria}.
    \item Identify the maximum ICARUS coverage in the ($m_X,\epsilon$) parameter space by evaluating the detector performance through the full detector simulation on the optimal mass parameter sets, $(m_0,m_1,m_2)$ determined in Step 1 above, scanning over the dark photon parameter space near the current exclusion limit.
\end{enumerate}

In the second step above, we find that the choice of $m_0$ mostly influences the optimal parameter set determination. Schematically, the expected number of events is proportional to the BDM flux and the fiducial cross-section satisfying all the selection criteria described in the previous section. As indicated by Eq.~\eqref{eqn:chiflux}, small $m_0$ values are preferred to enhance the BDM flux. On the other hand, $m_0$ sets the scale of visible energy deposits and the likelihood of meeting the displaced vertex requirement, i.e., selection criterion 2. Thus, the optimal $m_0$ value should, in principle, balance these considerations. For practical purposes, however, we select four benchmark mass points: $m_0=1$~GeV, 2 GeV, 5 GeV, and 10 GeV for our study.

Finally, the iBDM signal events for the parameter space determined through the steps above undergo a detailed full detector simulation, as described in Sec.~\ref{Simulation}, and their track and shower topology are used for the training of the scanners to identify the signal events in the real data. 
As mentioned earlier, an example of the wire signals and track topology expected from an iBDM interaction in the ICARUS detector is shown in Fig.~\ref{fig:iBDMeventdisplay}. 

\subsection{Simulation Study I: Accessible ($m_0$,$m_1$,$m_2$)} \label{simulationstudyI}
Assuming that all backgrounds can be vetoed through the application of the selection criteria and interaction vertex considerations, this analysis implements a zero background assumption. Therefore, the expression for the expected number of iBDM events, $N_{\rm expected}$ that satisfy all selection criteria and are captured in the scanning process, for a given detector exposure time, $t_{\rm exposure}$ is
\begin{equation}\label{eqn:expectedevents}
    N_{\rm expected} = N_e\;t_{\rm exposure} \; F_{\chi_1} \; \sigma_{\chi_1e^-\rightarrow \chi_2e^-}\times \xi_{\rm GE} \times \xi_{\rm scanning}
\end{equation}
\begin{equation}
\xi_{\rm GE}=\xi_{\rm criteria} \; \xi_{\rm trigger}\; \xi_{\rm filter}\\
\end{equation}

\begin{figure*}[t]
    \centering
    \includegraphics[width=0.47\linewidth, height=0.29\linewidth]{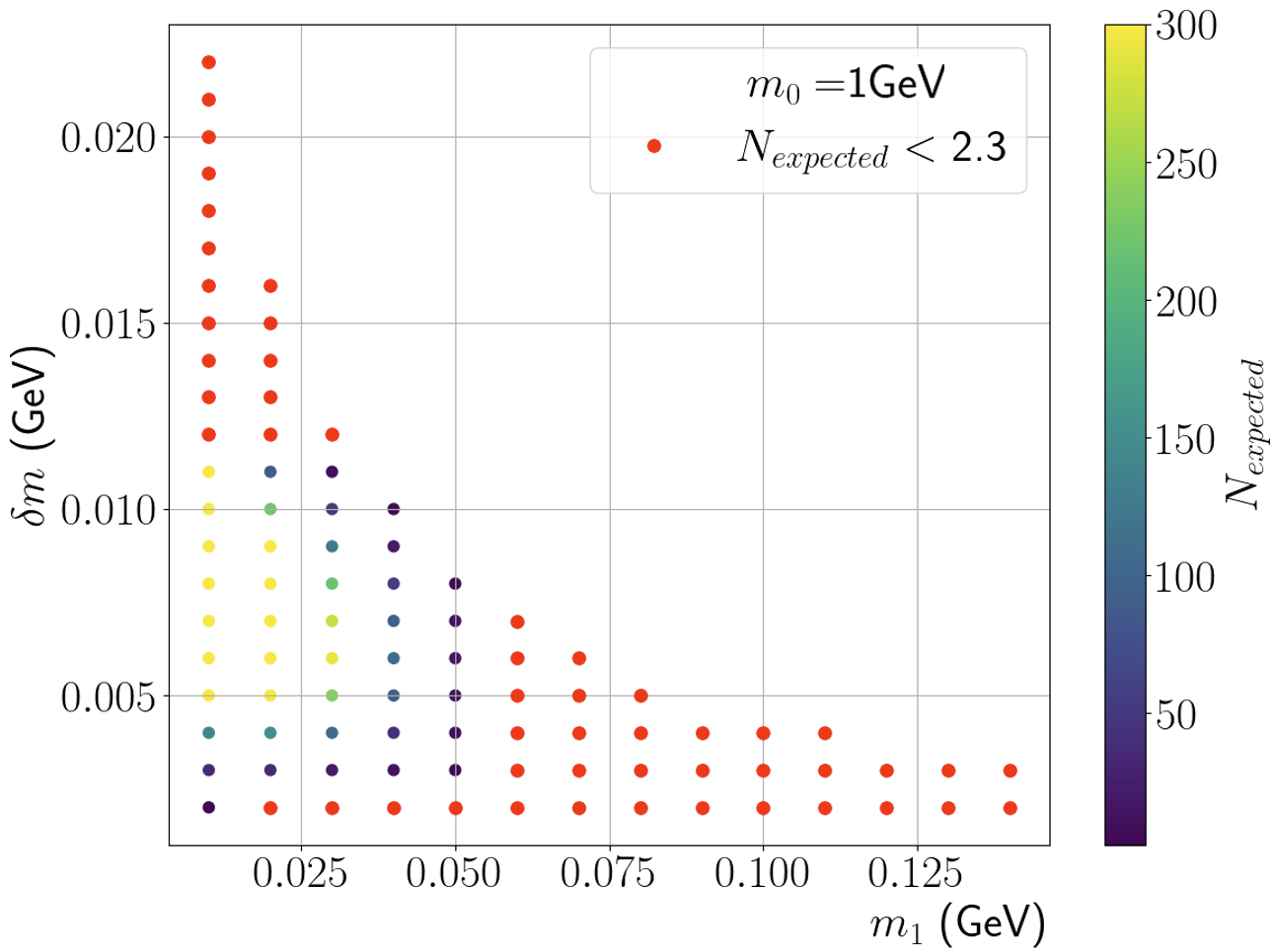}
    \includegraphics[width=0.48\linewidth, height=0.29\linewidth]{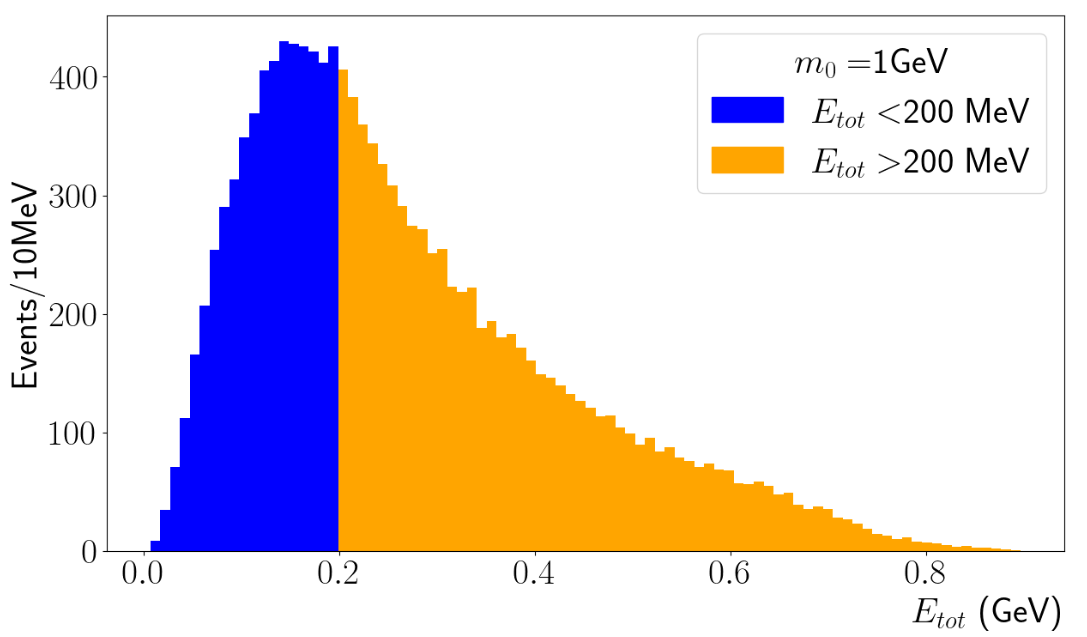}
    \includegraphics[width=0.47\linewidth, height=0.29\linewidth] {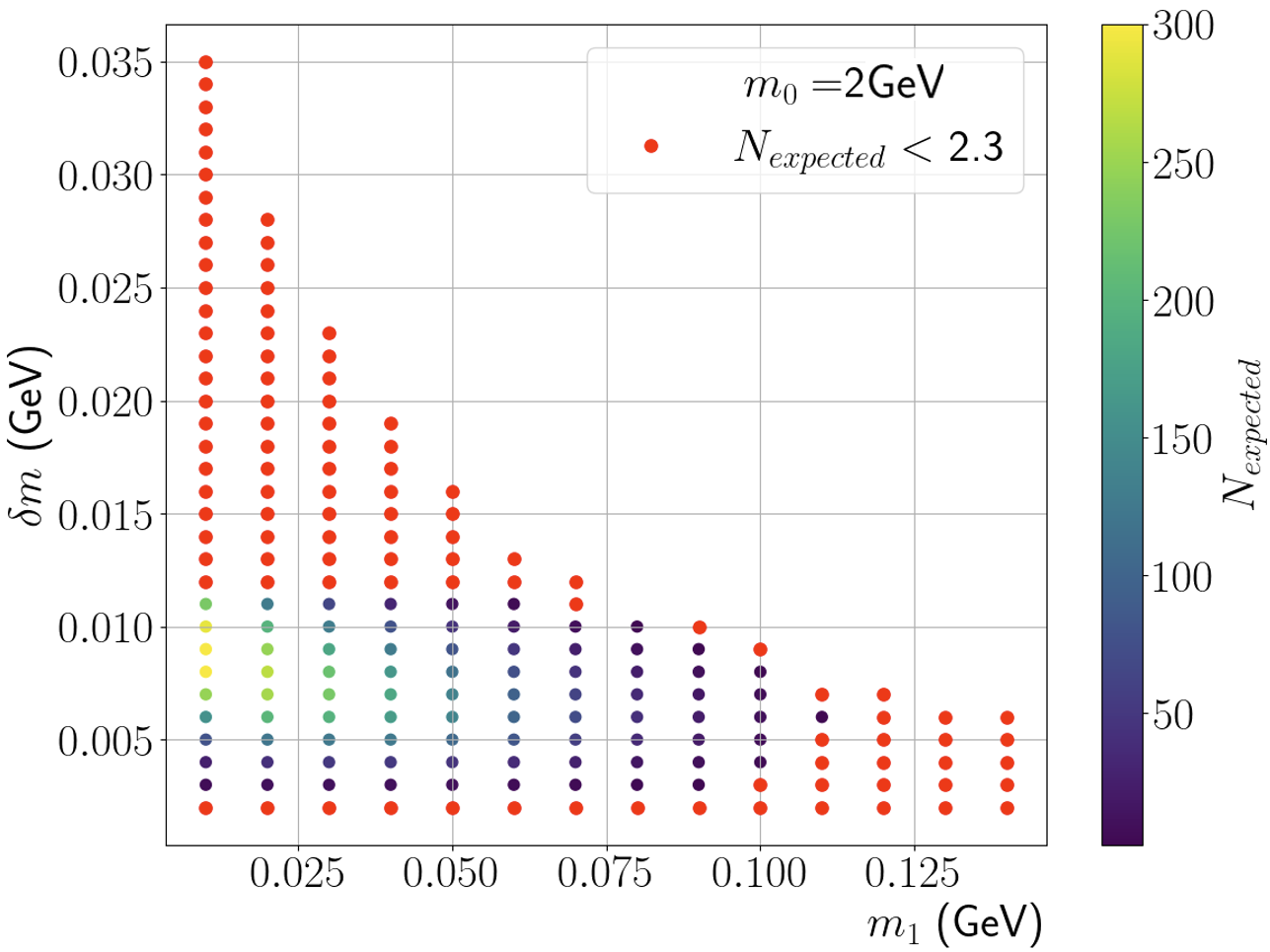}
     \includegraphics[width=0.48\linewidth, height=0.29\linewidth]{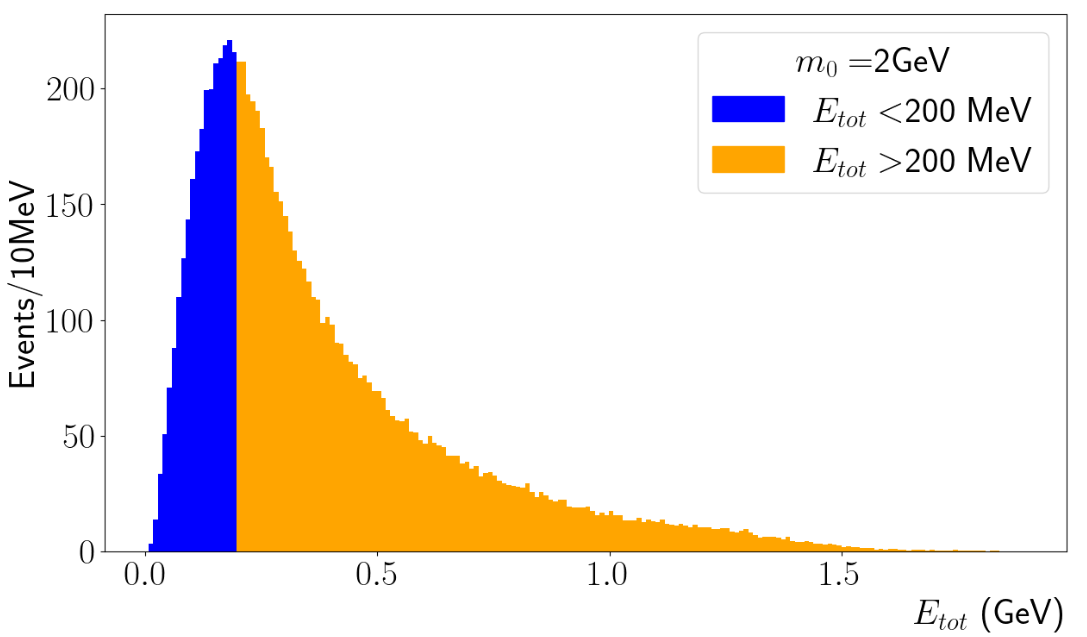}
     \includegraphics[width=0.47\linewidth, height=0.29\linewidth]{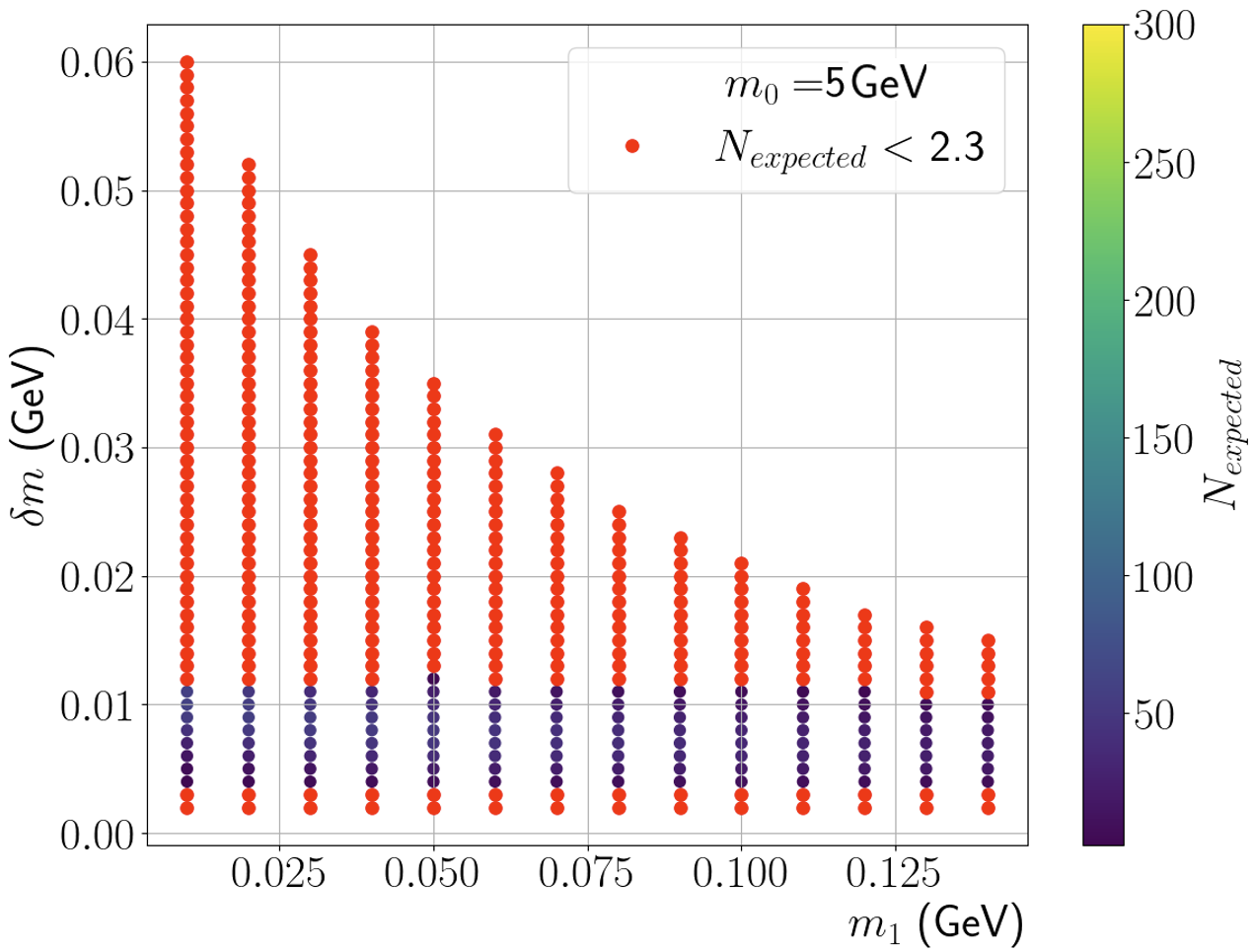}
     \includegraphics[width=0.48\linewidth, height=0.29\linewidth]{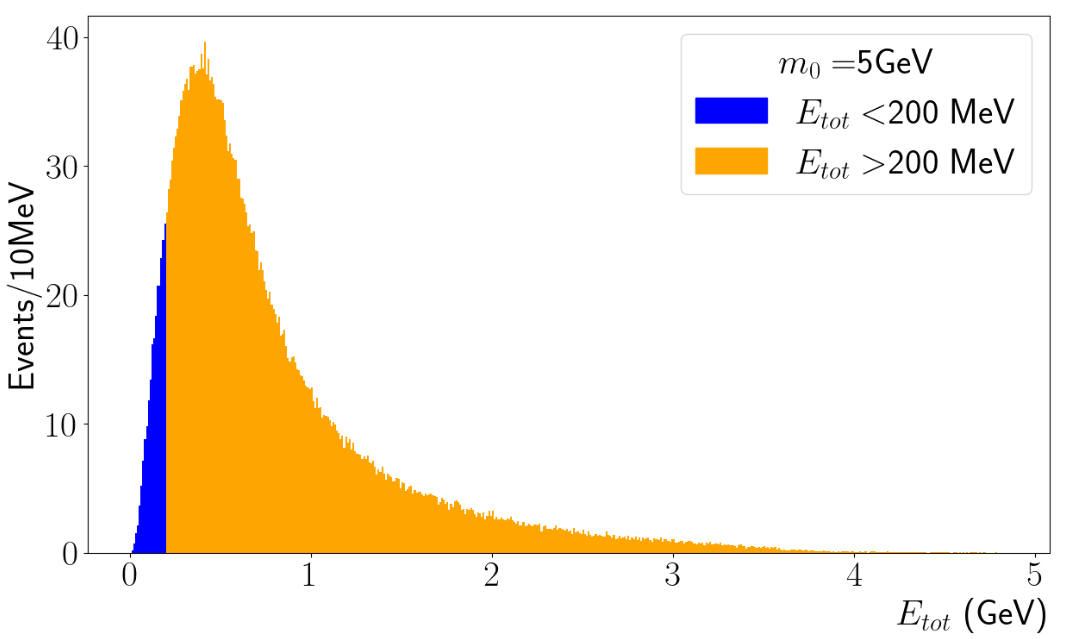}
    \includegraphics[width=0.47\linewidth, height=0.29\linewidth]{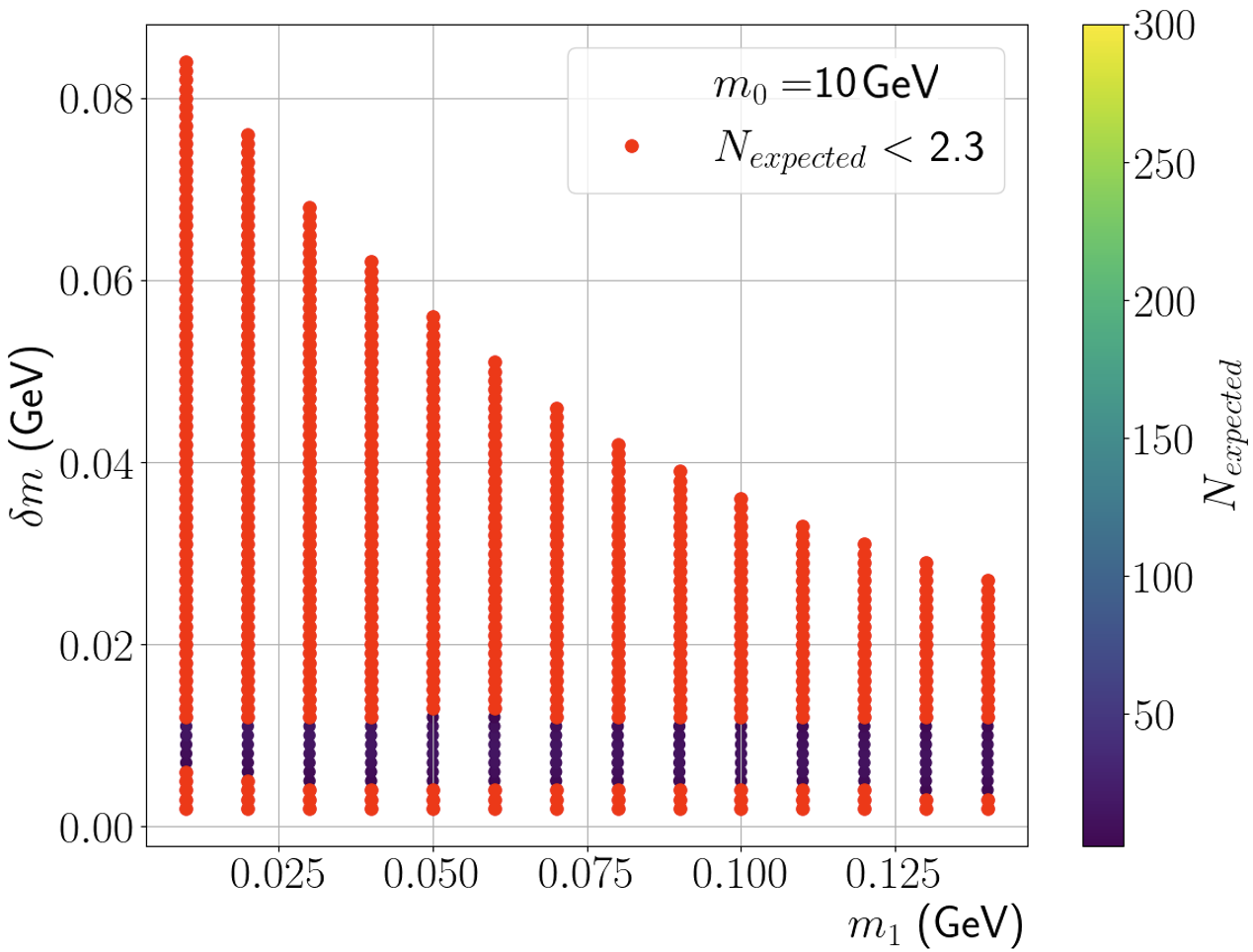}
    \includegraphics[width=0.48\linewidth, height=0.29\linewidth]{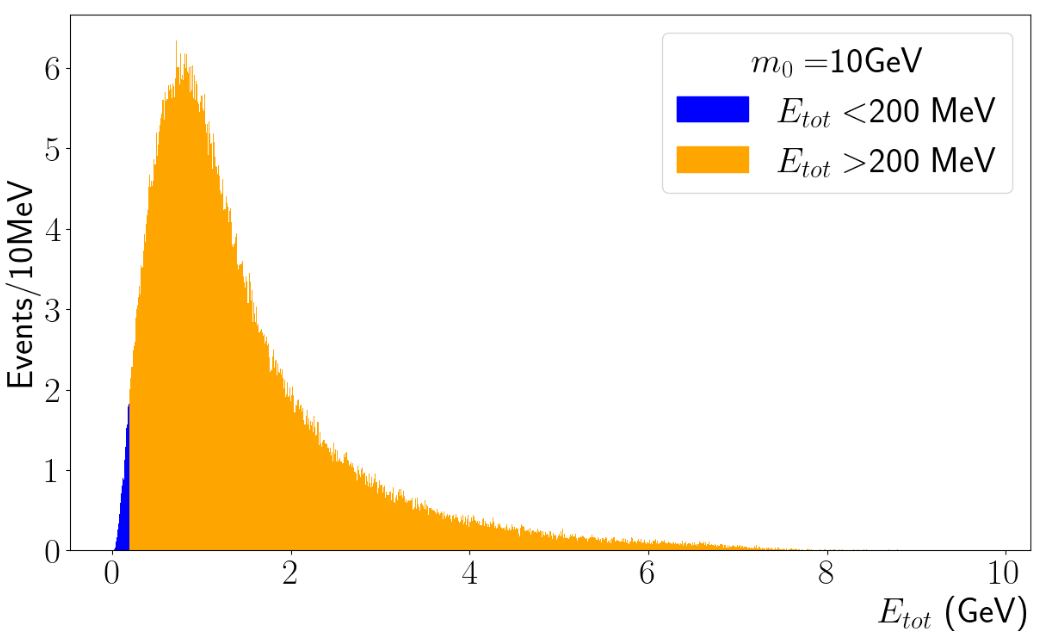}
    \caption{The plots on the left are the $(m_1, \delta m$) mass phase space with red points signifying $N_{\rm expected}<~$2.3 and colored points signifying $N_{\rm expected}>~$2.3, where the color grading represents $N_{\rm expected}$. The $N_{\rm expected}$ value is saturated to 300 to show ($m_1,\delta m$) points where $N_{\rm expected}$ is maximum. Here, the points in red are categorized as inaccessible under the 90\%~C.L. assumption with the selection criteria imposed. The color distribution indicates the number of events expected for $t_{\rm exposure}=0.3~{\rm year}$ that pass the selection criteria presented in Sec~\ref{selectioncriteria}. Recoil electron + electron-positron pair total energy distribution (right) for points $N_{\rm expected}>~$2.3 (all non-red points) normalized to the total number of expected events for all non-red points $N_{\rm tot}=\sum_{\rm points} N_{\rm expected}$. }
    \label{fig:massspan}
\end{figure*}
where $N_e$ is the number of electrons in the fiducial volume, $F_{\chi_1}$ is the $\chi_1$ flux defined in Eq.~\eqref{eqn:chiflux}, $t_{\rm exposure}$ is the detector's total exposure time, and $\xi_{\rm GE}$ is the global efficiency. The exposure time for the analyzed data set 0.13~kton$\cdot$year is $\sim0.3$~years.
The scanning efficiency $\xi_{\rm scanning}$ quantifies the ability of a scanner to identify tracks in the data that are topologically similar to the iBDM candidates studied within a sample of simulated events such as that shown in Fig.~\ref{fig:iBDMeventdisplay}.

For the analysis presented in this section, detector, and filter algorithms will be forgone until the detector simulation stage after obtaining optimal mass parameter sets to cover as much space in the ($m_X,\epsilon$) available parameter space. This implies that this section implements the selection criteria on the Monte Carlo generated and uses the number of events that pass the selection criteria as a metric, assuming 100\% for the filter efficiency, trigger efficiency, and scanning efficiency. The procedure below is followed to identify optimal ($m_0,m_1,m_2$) mass sets at the edge of the dark photon parameter space ($m_X,\epsilon$)$_{\rm limit}=$~(12~MeV,~8$\times$10$^{-4}$): 

\begin{enumerate}
    \item Assuming a fixed $m_0$ with value $\mathcal{O}(1-10~{\rm GeV})$, impose $0.010~{\rm GeV}\leq m_1 \leq 0.150~{\rm GeV}$ and span with a granularity of $\Delta m_1$=~10~MeV.
    \item Make kinematically allowed~\cite{indirectDMdetection_2014,De_Roeck_2020} $(m_1,m_2)$ pairs, with a $m_2$ granularity of $\Delta m_2$~=~1~MeV.
    \item Generate iBDM events and apply selection criteria to events to obtain $\xi_{\rm criteria}$.
    \item Identify all ($m_0,m_1,m_2$) mass sets that have $N_{\rm expected}\geq$~2.3 and identify where $N_{\rm expected}$ is maximum in the $(m_0,m_1,m_2)$ space at ($m_X,\epsilon$)$_{\rm limit}$ to obtain the optimal mass parameter sets for a given $m_0$ assuming $\xi_{\rm filter}=$~1 and $\xi_{\rm trigger}=$~1.
\end{enumerate}

Figure~\ref{fig:massspan} shows the results of this analysis. Each row contains two plots for $m_0=1$~GeV, 2~GeV, 5~GeV, and 10~GeV, as indicated on the plot. The plots on the left column show $\delta m$ vs $m_1$, where $\delta m=m_2-m_1$. The red ($m_1,\delta m$) points on the plots are inaccessible to ICARUS, in which the expected number of events $N_{\rm expected}$ after all selection criteria are applied is less than 2.3, the limit that fails to satisfy the 90\% C.L. limit at the present dark photon exclusion limit, ($m_X, \epsilon$)$_{\rm limit}$. All other colored points in these plots are the accessible mass parameter sets to ICARUS in which $N_{\rm expected}\geq$~2.3, assuming $\xi_{\rm trigger}$, $\xi_{\rm filter}$ and $\xi_{\rm scanning}$ are all equal to 1. The color scale of the non-red points indicates the value for $N_{\rm expected}$. The color scale is saturated at 300 on purpose to allow points with a smaller number of $N_{\rm expected}$ can be visible.  In other words, all combinations of masses with $N_{\rm expected}$ above 300 on Table~\ref{table:1} which lists a few optimal $(m_0, m_1, m_2)$ sets which result in rather large number of $N_{\rm expected}$ are all in yellow on the plots.

\begin{table}[b]
\begin{tabular}{lllll}
\toprule
 $m_0$ (GeV) & $m_1$ (MeV) & $m_2$ (MeV) & $N_{\rm expected}$\\
 \midrule
 1 & 10 & 18 & 940\\
 1 & 20 & 26 & 720\\
 1 & 30 & 36 & 511\\
 1 & 50 & 55 & 231\\
 2 & 10 & 19 & 313\\
 2 & 20 & 28 & 278\\
 2 & 30 & 37 & 249\\
 2 & 50 & 56 & 182\\
 5 & 10 & 20 & 61\\
 5 & 20 & 29 & 58\\
 5 & 30 & 39 & 55\\
 5 & 50 & 58 & 51\\ \bottomrule
 \hline
\end{tabular}
\caption{The list of optimal DM mass parameter sets for which $N_{\rm expected}$ is maximum for ($m_X,\epsilon)=($12~MeV, 0.0008), the present exclusion limit of the dark photon parameter space. Selection criteria are imposed on events; hence, $\xi_{\rm criteria}$ is applied, whereas trigger, filter, and scanning efficiencies are assumed 100\%.}
\label{table:1}
\end{table}

The plots on the right column show the total energy $E_{\rm tot}$ of the visible, outgoing particles from the iBDM primary ($e_R^-$) and secondary ($e^-e^+$) interactions. 
The boundary between the orange and blue shaded areas indicates the 200~MeV energy threshold.
The plots show that as $m_0$ increases, the total energy of the visible particles in the detector increases, and the fraction above the 200~MeV threshold increases. It, however, is clearly seen on the vertical scale of the plots that the number of expected events, $N_{\rm expected}$ passing all other criteria decreases as $m_0$ increases. This is due to the fact that the $F_{\chi_1}$ is inversely proportional to the $m_0$ mass squared, therefore, $N_{\rm expected}$ is also scaling as $1/m_0^2$. This is also visible in corresponding plots on the left where the scale of non-red points decreases as $m_0$ increases.

For every $m_1$, there is a $(m_1, \delta m)$ mass pair that maximizes $N_{\rm expected}$, as is apparent in the $m_0=1$~GeV and $m_0=2$~GeV cases (maximally yellow point), as can also be seen in Table~\ref{table:1}. 
While $m_0 = 1$~GeV has the larger $N_{\rm expected}$, the energy distribution peaks closer to $E_{\rm thres}$, reducing the chances that the 90\% C.L. criteria to be satisfied once the selection criteria are applied, and all efficiencies are evaluated with the full detector simulation.

Focusing on $m_0$ = 1 GeV and 2 GeV, the exclusion of some combinations can be easily traced to the adopted selection criteria. The $m_2$ masses in which $\delta m > m_X $ create on-shell dark photons, producing iBDM interactions with prompt $\chi_2$ decay and a subsequent prompt $X$ decay, with the average decay lengths $< 1$ cm for ($m_X,\epsilon$) = (12 MeV, 0.0008)~\cite{De_Roeck_2020}. This condition makes the events fail the 3 cm minimum distance selection criteria between the primary and secondary vertices.  

Alternatively, the whole bottom row of red points for small $\delta m$ has a large fraction of events with $\chi_2$'s with long lifetimes, which fail to be selected, since they likely are decaying outside the fiducial volume~\cite{De_Roeck_2020}.  Lastly, for a large $m_1$, a significant fraction of the kinetic energy is used for the $m_1$ and $m_2$ masses, and although many events could be above the threshold energy $E_{\rm thres}$, the number of expected events are too small to satisfy the 90\%~C.L. ($N_{\rm expected}<2.3$).

For $m_0$ = 5 GeV and 10 GeV, the energy threshold affects little on the large mass regions of ($m_1,\delta m$) due to the sufficient energy supplied to $\chi_1$ as is seen in the energy spectrum for the respective $m_0$ masses. However, both the on-shell dark photon effect ($m_X < \delta m$) which causes the events to fail the 3~cm minimum primary-secondary vertex separation requirement and for the small $\delta m$ region, a large fraction of events fail the full event fiducial volume containment requirements create the boundaries to the blue and red points.
Thus, these mass points are inaccessible to ICARUS.


The energy distributions in Fig.~\ref{fig:massspan} show that increasing $m_0$ decreases the number of expected events overall, reducing the dark photon parameter space coverage by the ICARUS detector.  Conversely, the energy range increases as $m_0$ increases, enabling more energetic interactions in the detector. Since only $\xi_{\rm criteria}$ is applied to these events, $\xi_{\rm filter}$ and $\xi_{\rm trigger}$ still need to be considered. For $m_0=$~1~GeV and $m_0=$~2~GeV, the number of events per 10~MeV energy bin is significantly higher than $m_0=$~5~GeV and $m_0=$~10~GeV. This behavior is mainly attributed to the $\chi_1$ flux and $m_0$ inverse relationship as in Eq.~\eqref{eqn:chiflux}.

Although $m_0=1$~GeV has a greater $N_{\rm expected}$ than a similar optimal mass set with $m_0=2$~GeV as can be seen in Table~\ref{table:1}, the total energy distribution of the recoil electron and electron-positron pair, $E_{\rm tot}$ for $m_0=$~1~GeV has a peak at slightly above 100 MeV---which is much below $E_{\rm thres}=200$~MeV---with the maximum energies for all the accessible ($m_1,\delta m$) mass combinations stay below 800~MeV, as can be seen in Fig.~\ref{fig:massspan} top right. More quantitatively, about 55\% of events are above the energy threshold. On the other hand, the $E_{\rm tot}$ distribution for $m_0=$~2~GeV peaks near the threshold, resulting in approximately 67\% of events exceeding the threshold. Moreover, the energy distribution spans to ~1.5~GeV, increasing the likelihood of satisfying the displaced vertex requirement, i.e., selection criterion 2. Consequently, the extended energy distribution to higher energies enables the ($m_1,\delta m$) mass sets of $m_0=$~2~GeV to have more events in the higher trigger efficiency range.

All in all, both $m_0=$~1~GeV and $m_0=$~2~GeV enable the search of iBDM in the unexplored ($m_X, \epsilon$) parameter space, making these masses optimal $m_0$ values. In addition, Table~\ref{table:1} shows which $m_2$ makes $N_{\rm expected}$ maximum for a given ($m_0,m_1$) mass pair. The number of events is significantly greater for all ($m_1,m_2$) mass pairs for $m_0=$~1~GeV and $m_0=$~2~GeV. Due to the $\chi_1$ flux factor having an inverse relationship with $m_0$ [see Eq.~\eqref{eqn:chiflux}], the number of events for $m_0=$~5~GeV and $m_0=$~10~GeV are significantly less, therefore when the filter and trigger efficiencies are applied, the parameter space span will be significantly reduced.

Given the number of events for both $m_0=$~1~GeV and $m_0=$~2~GeV, if there are any performance improvements at the level of the filter algorithm, it is minimal compared to the impact of the selection criteria efficiency $\xi_{\rm criteria}$ due to the lower energies at $m_0=$~1~GeV. The dark photon mass $m_X$ has a significant impact on the lifetime of $\chi_2$, therefore affecting both the FV vertex containment criteria and 3~cm primary-secondary distance criteria. The $m_0=$~2~GeV extended energy distribution in the tail enables the primary-secondary vertex separation criterion to be respected by more events every $m_X$ while not allowing an over extension to also respect the FV containment requirement. These observations motivate us to choose $m_0=$~2~GeV as the reference parameter for this analysis.

\subsection{Simulation Study II: $\xi_{\rm GE}$ and $\epsilon$ vs $m_X$}\label{simulationstudyII}
The optimal mass parameter sets identified in the previous section are used for the remainder of this study, namely $m_0=2$~GeV and $(m_1, m_2)$ = (10 MeV, 19 MeV), (20 MeV, 28 MeV), {\rm and} (30 MeV, 37 MeV). For these parameters combinations, the yield $N_{\rm expected}$ is higher than for $m_0> 2$~GeV, and the total energy deposited in the detector is higher than that for  $m_0= 1$~GeV. This section presents the results from a full detector simulation for these mass sets for sample points at the present exclusion limit of the dark photon ($m_X,\epsilon$) parameter space and obtain the global efficiency $\xi_{\rm GE}=\xi_{\rm criteria} \; \xi_{\rm filter} \; \xi_{\rm trigger}$. 

The efficiency for the selection criteria, $\xi_{\rm criteria}$ is determined as described in Sec.~\ref{simulationstudyI}. 
The maximum reachable dark photon parameter space is determined based on the most optimal global efficiency, $\xi_{\rm GE}$, which depends on the efficiency for passing the selection criteria, $\xi_{\rm criteria}$.
The procedure to obtain $\xi_{\rm GE}$ for each dark photon parameter for the given optimal dark sector mass set is as follows: 
\begin{enumerate}
    \item Simulate 5,000 events for each $(\epsilon, m_X)$ sample space point under the given mass set ($m_0,m_1,m_2$). This simulation sample size is chosen to optimize the analysis process, while maintaining the statistical uncertainty below 10\%, achieving as low as 2\% uncertainty for the global efficiency $\xi_{\rm GE}$. 
    \item Perform the detailed GEANT4 detector simulation of the recoil electron and associated electron-positron pair interaction with LAr in the detector for all 5,000 events to obtain the ionization charge information and the total deposited energy.
    \item Use the GEANT4 output as the input for the wire simulation programs used for ICARUS at Gran Sasso. This step automatically applies the filter algorithm and evaluates the filter efficiency $\xi_{\rm filter}$ and the trigger efficiency $\xi_{\rm trigger}$.
\end{enumerate}

\begin{figure}[t]
    \centering
    \includegraphics[width=1\linewidth]{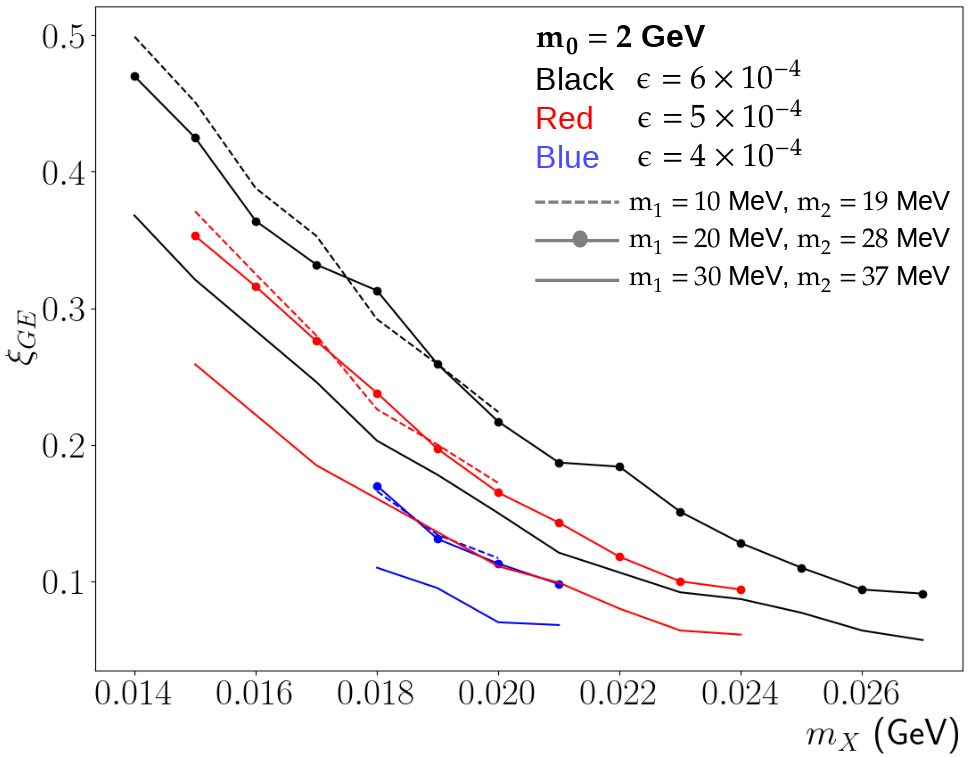}
    \hfill
    \caption{Plot showing the global efficiency $\xi_{\rm GE}$ as a function of $m_X$ for various $\epsilon$ (line color) and three optimal ($m_1,m_2$) mass sets (line type) for $m_0=~$2~GeV. This graph shows that an increase in $\epsilon$ and/or a decrease in $m_X$ has a positive effect on our ability to trigger and filter events that satisfy the selection criteria presented in Sec.~\ref{selectioncriteria}.}
    \label{fig:GEplot}
\end{figure}

Figure~\ref{fig:GEplot} shows the global efficiency, $\xi_{\rm GE}$, as a function of the dark photon mass $m_X$ for the given mass parameter sets ($m_0, m_1, m_2$)  and the mixing parameters, specified on the plot. The different colors are the different mixing parameter values, and the different line types represent the three different mass parameter sets. 

The global efficiency, $\xi_{\rm GE}$, decreases as the dark photon mass $m_X$ increases or as the kinetic mixing parameter $\epsilon$ decreases. These trends are due to an interplay between the fiducial volume containment requirement and the minimum energy deposit requirement.  
When the dark photon mass $m_X$ increases or the kinetic mixing parameter $\epsilon$ decreases, the lifetime of the unstable dark sector state $\chi_2$ increases, and the distance between the primary and secondary vertices also increases. As a consequence, the probability that the secondary vertex falls outside the fiducial volume also increases, and the total energy deposited in the detector decreases due to the secondary particles exiting the detector.

\subsection{Real Data Analysis}\label{realdataanalysis}
As described in Sec.~\ref{The ICARUS Detector}, the data specifically filtered for the atmospheric neutrino study required the presence of one or more e.m. showers with the total energy deposit in the event above 200~MeV.
Since the iBDM signature for this analysis consists of an e.m. shower of the electron from the primary interaction followed by an associated e.m. shower of the $e^{+}e^{-}$ pair from the secondary interaction, the preselected data set for the atmospheric neutrino study is expected to contain iBDM candidates. 
The ICARUS data set used for this iBDM search is about 30\% of the data recorded in the 2012--2013 ICARUS operational run, corresponding to an exposure of 0.13~kton$\cdot$year. 
In order to further select the iBDM candidates, a total of 4,134 filtered events in this data set are scanned visually.
The scanning criteria for selecting an iBDM candidate event starts with requiring e.m. showers clearly separated from a nearby track which could be that of a cosmic ray muon.
Each event is visually inspected using the event displays and the three wire plane views, as in the atmospheric neutrino study~\cite{universe501001}. 
The scanned events are classified into the following four categories for ease of follow-up analysis:
\begin{enumerate}
    \item Event with only noisy wires
    \item Event with an identified muon with no isolated showers
    \item Event with a vertex from which multiple tracks emerge
    \item Event with an isolated shower
    \item Event that requires further investigation.
\end{enumerate}

In order to take as conservative an approach as possible, a preliminary visual scan does not reject muons that have showers that appear different from delta rays, which in general point back to the track from which they are emitted. 
In the subsequent detailed analysis, the events classified as categories 1, 2, and 3 are rejected and excluded from the further investigation. 
The events in categories 4 and 5 are subject to a subsequent detailed investigation. 
Any events in category 4 with isolated showers are rejected if they contain an identifiable muon track.

Once the scanning and the categorization are complete, the $dE/dx$ track identification is performed on the final set of the remaining four events that survived the entire selection criteria, including the detailed inspection above. This process verifies that the primary track is an electron, satisfying the m.i.p. signature described in Ref.~\cite{universe501001}, as well as the associated secondary track having the two-m.i.p. signature, the indication of an electron-positron pair.

\subsection{Evaluation of Uncertainties}\label{EvaluationofUncertainties}

Various uncertainties of the ICARUS detector performances, such as the spatial resolution and the energy resolution, could directly impact the selection of iBDM events. Therefore, the event selection criteria that are dependent on performance parameters must be taken into account in estimating systematic uncertainties in the iBDM search, in addition to the efficiencies for the filter, the trigger, and the event scanning.

The spatial resolution of the detector is measured to be $\sim 1~{\rm mm}^{3}$ as described in Ref.~\cite{ICARUSbible,ICARUSTrackReconstructionPaper}. This uncertainty directly impacts the number of potential candidates since the selection criteria require full containment of both the primary and secondary vertices in an iBDM event within the fiducial volume, defined as 5~cm inward from the boundaries of the active volume, and minimum 3~cm distance separation between the primary and the associated secondary vertices. 
The fractional uncertainties of the spatial resolution are $\sim3\%$ to the minimum distance requirement and $\sim2\%$ to the fiducial volume criterion.
The impact of the spatial resolution to the overall uncertainty of the selection efficiency, $\xi_{\rm criteria}$ is estimated using the standard technique of varying the cut value for each requirement by $\pm 1$~mm in all directions and taking the fractional differences of the number of expected events that pass between the varied cut values.
The resulting percentage uncertainty due to spatial resolution to the fiducial volume containment requirement is estimated to be $<0.1\%$.
On the other hand, the 3~cm minimum distance requirement uncertainty is estimated to $+2\%$ and $-1\%$ due to the exponentially falling spectrum of the distance between the two vertices. We take 2\% as the uncertainty for this requirement to be conservative.
The resulting combined percentage systematic uncertainty for selection efficiency, $\xi_{\rm criteria}$ for both fiducial volume and the 3~cm minimum distance requirements, due to the detector spatial resolution is $\pm 2\%$. 

The energy resolution for the reconstructed e.m. shower was evaluated
to be $\sigma/E({\rm GeV})=3\%/\sqrt{E} \bigoplus 1\%$~\cite{ICARUSinitial} by the ICARUS collaboration, studying the reconstruction of the $\pi^0$ events. This resolution corresponds to a few percent at the typical energy of the events considered for the iBDM interactions (around 1 GeV). 
The energy resolution impacts directly the present analysis since we implement a 200 MeV minimum energy deposition requirement to which the impact of the energy resolution at this threshold is at the level of $7.7\%$. As in the case of position resolution, the standard technique of varying the energy threshold by $\pm7.7\%$ is applied to the energy resolution uncertainty estimate, resulting in the percentage uncertainty $\delta N_{\rm expected}/N_{\rm expected}\sim 1\%$.

The overall systematic uncertainty due to the selection criteria is estimated by adding the three uncertainties above in quadrature, resulting in $\pm 2.2\%$.

The filter efficiency, $\xi_{\rm filter}$ across the dark photon sample space is found to be $\sim 93\%$ on average, estimated by applying the filter criteria to the detailed signal simulation sample. 
Its percentage statistical uncertainty, $\delta \xi_{\rm filter}/\xi_{\rm filter}$ is found to range 1.7\% -- 2.3\%, depending on the DM model parameter sets.
The resulting systematic uncertainty due to the filter efficiency to present analysis is obtained using the same methodology as the above, namely varying the efficiency by the corresponding uncertainty, estimated to range 0.8\% -- 1.1\%.
We take 1.1\% systematic uncertainty due to the statistical uncertainty of the filter efficiency to be conservative.
In addition, due to the fluctuation in the filter efficiencies across the different parameter space, we reflect 1.5\% additional systematic uncertainty due to the filter efficiency.
The overall systematic uncertainty due to the filter efficiency, therefore, is estimated as the quadratic sum of the two uncertainties, resulting in 1.9\%.

Similarly, the percentage uncertainty of the trigger efficiency determined by the ICARUS collaboration, $\delta \xi_{\rm trigger}/\xi_{\rm trigger}$ is estimated to range between 1\% and 2\%, resulting in 1.5\% systematics to this analysis.

These uncertainties are incorporated into the uncertainty on the global efficiency $\xi_{\rm GE}$ since it includes the efficiencies of the trigger, selection criteria, and filter algorithm.

Finally, the scanning efficiency is estimated using a blind set of detailed simulations of the iBDM signal events.
Each scanner is asked to scan the blind set of iBDM signal sample and categorize the events as described in Sec.~\ref{realdataanalysis}.
The efficiency for the event categorization is found to be consistent between the scanners, resulting in the overall efficiency of $\xi_{\rm scanning}=76\%\pm 5\%$, where the uncertainty is estimated by adding the statistical uncertainties of the scanner efficiencies in quadrature to be conservative.

Table~\ref{table:uncertainties} summarizes the sources of the uncertainties and their contributions to the overall uncertainties in this analysis, which is reflected in the final results.

\begin{table}[t!]  
  \centering
  \begin{tabular}{lll}
    \toprule
    Source &  Uncertainty \\ \midrule
    $\xi_{\rm criteria}$ &  2.2\% \\
    $\xi_{\rm filter}$ & 1.9\% \\ 
    $\xi_{\rm trigger}$ & 1.5\% \\ 
    $\xi_{\rm scanning}$ & 5\% \\ \hline\hline
     Total Uncertainty & 6\% \\ \bottomrule

    \bottomrule
  \end{tabular}
  \caption{The total systematic uncertainty from each source of the uncertainties. These values are reflected in the 1$\sigma$ deviation of the dark photon exclusion limit.}
  \label{table:uncertainties}
\end{table}

\section{Background Estimate}\label{Backgrounds}
From the atmospheric neutrino analysis, $\nu_e$CC events were identified for the analysis presented in Ref.~\cite{universe501001}. The same filtered dataset used to identify $\nu_e$ events is used for the iBDM search. 
Cosmic ray muons and neutrino interactions may produce e.m. showers in the detector that could mimic the iBDM signal. 
In this section, the estimate and the rejection strategy of the backgrounds from various sources, including these, are presented.

\subsection{Cosmic Ray Muon Background}
\begin{figure}[t]
    \includegraphics[width=1.0\linewidth]{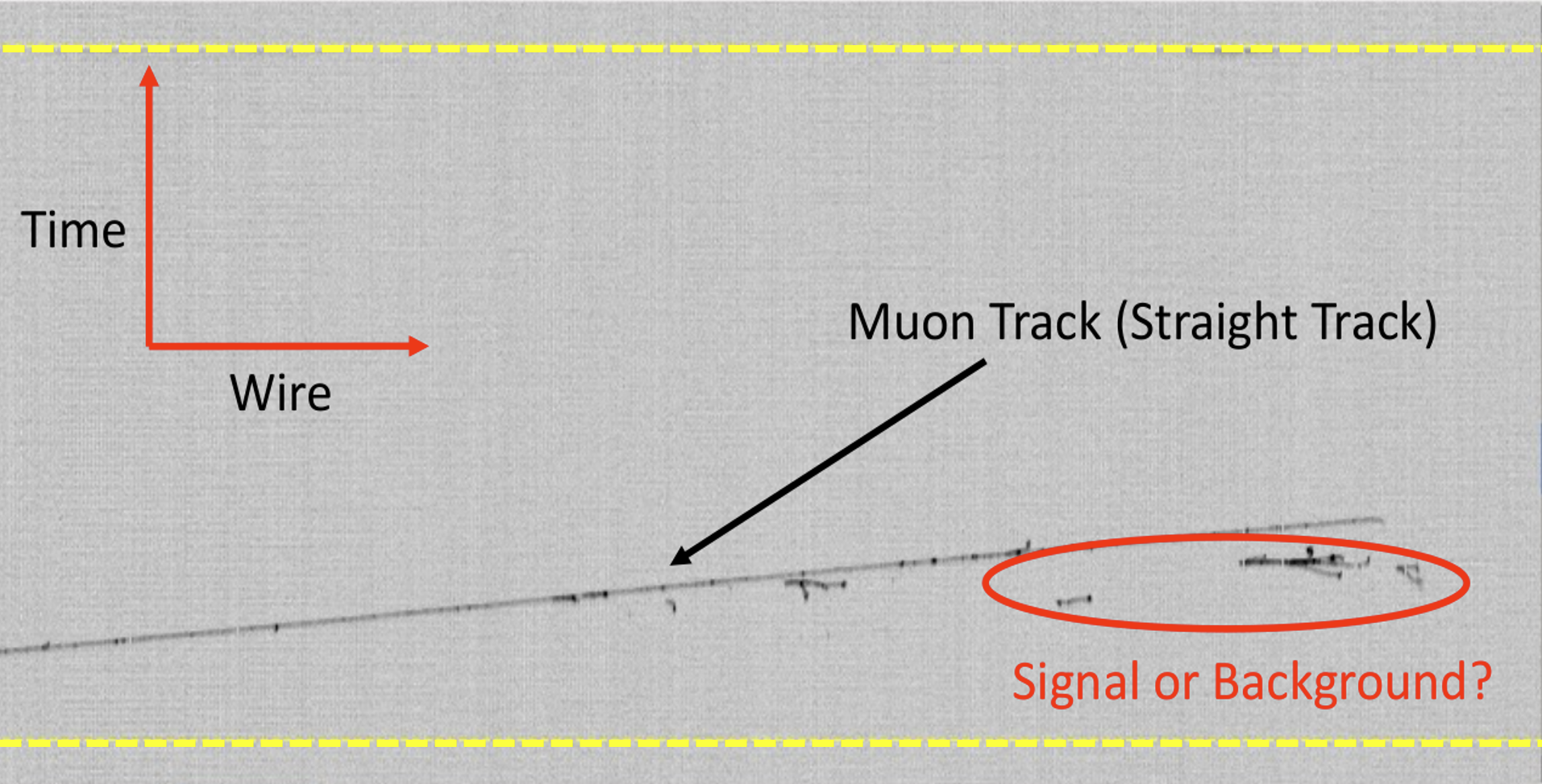}
    \hfill
    \caption{An example collection plane view of e.m. activity produced by a through-going cosmic muon that spans about 1.2~m horizontally, recorded by the detector. The two yellow dashed lines indicate the anode and the cathode planes at 1.5~m distance vertically.
    Depending on the deposited energy, the showers produced by the delta rays could become a source of the background for both the recoil electron and electron-positron pair in the iBDM signal.}
    \label{fig:muoneventdisplay}
\end{figure}

Despite the $\sim 10^6$ reduction of the cosmic ray flux by the 3,400 m.w.e. overburden at LNGS, high-energy cosmic ray muons could still reach the ICARUS detector and produce e.m. showers through delta rays and emissions of bremsstrahlung photons.
As shown in Fig.~\ref{fig:muoneventdisplay}, delta rays are attached to the crossing muon while bremsstrahlung photons and secondary photons from delta rays can generate e.m. showers sufficiently isolated from the muon track, as the tracks circled in red. These isolated e.m. showers could have energies above the threshold and mimic the iBDM signal when they are separated from their accompanying and parent muon, which may or may not enter the active volume of the detector.

There is also the possibility of a low energy muon entering the detector's sensitive volume and ranging out, mimicking the track of a non-showering m.i.p.-like signal similar to an electron~\cite{iBDMprotodune}. The fiducial volume requirement, which cuts 5~cm into the active volume, helps reject these backgrounds since their track extends outside of the fiducial volume.

\begin{figure}[t]
    \centering
    \includegraphics[width=1.0\linewidth]{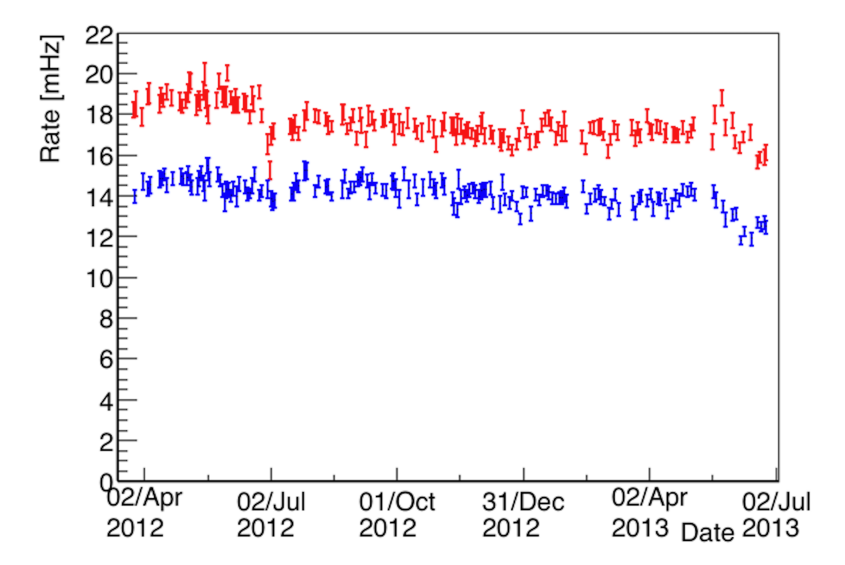}
    \hfill
    \caption{The muon rate as measured in the $1^{\rm st}$ (blue points) and $2^{\rm nd}$ (red points) ICARUS modules over the 2012--2013 ICARUS operational run~\cite{universe501001}. Only statistical uncertainties are shown. Some variations visible along the run are related to the detector conditions, including LAr purity and presence of the noise impacting the track reconstruction.}
    \label{fig:muonfluxrate}
\end{figure}

Figure~\ref{fig:muonfluxrate} shows the measured rate of the reconstructed cosmic muon tracks, arriving at the ICARUS detector in LNGS as a function of time. 
The considered data sample was collected in the  the April--November 2012 period. 
The rates are different in the two modules due to the difference in the number of PMTs in each of them, as described in Sec.~\ref{The ICARUS Detector}. Some variations visible along the run are related to the detector conditions, including LAr purity and the presence of noise that impacts the track reconstruction. Taking the average of the measured rates from both cryostats, the total muon rate is found to be $\Gamma \sim 32 \times 10^{-3}$~Hz~\cite{universe501001}. 

To effectively suppress muon-associated backgrounds, a conservative approach is taken, rejecting events with the presence of an identifiable muon, identified by a straight track.
In fact, the probability of rejecting an event  due to an uncorrelated cosmic muon randomly overlapping the readout window $P(\mu\vert t_{\rm drift})$ is proportional to the muon rate above and the  $\sim $1~ms readout window is : 
\begin{equation}
    N(\mu\vert t_{\rm drift}) \sim \Gamma t_{\rm drift} \sim 3.2\times 10^{-5}
\end{equation}
Therefore, this background is very effectively suppressed  with a negligible loss of acceptance by rejecting all events with an identifiable muon. 

\subsection{Atmospheric Neutrino Background}
\label{AtmosphericNeutrinoBck}

A study devoted to the search for atmospheric neutrino interactions in ICARUS at LNGS \cite{universe501001} permitted the identification of a small number of  $\nu_e$CC and $\nu_\mu$CC event candidates in exposure over 3 times larger (0.43~kton$\cdot$year) than the one analyzed in this paper (0.13~kton$\cdot$year). 
Neutrino interactions can produce e.m. activity by several mechanisms, such as the e.m. shower initiated by the primary electron in $\nu_e$CC interactions, delta rays or bremsstrahlung photons by muons in $\nu_{\mu}$CC interactions, and photons from $\pi^0$ decays. 

Based on the ICARUS atmospheric neutrino analysis study presented in Table~\ref{table:tableneutrinostudy}, a total of 1.3 $\nu_{\mu}$CC, 4 $\nu_e$CC, and 0.4 NC interactions from atmospheric neutrinos are expected to be contained in the data sample considered in the present analysis.
To be conservative, we estimate at the maximum 6 atmospheric neutrino backgrounds, mostly from $\nu_e$CC interactions in the sample.
These events, however, have signatures distinct from iBDM events that can be exploited to efficiently reject them.
In addition to the presence of hadronic activity at the neutrino interaction vertex, several signatures can be exploited to identify and distinguish neutrino interactions from the iBDM events: the observation of the primary muon in $\nu_{\mu}$CC interactions, the measurement of $dE/dx$ at the beginning of the showers together with the observation of the second shower in $\pi^0 \rightarrow \gamma \gamma$ \cite{ICARUSpaperLSND,ICARUSpaper}.

In addition, the identified events from the preliminary scan are compared to the atmospheric neutrino scanning results and removed from the final detailed inspection if they are already identified as such, thereby further reducing the potential background from this source. 
Applying the characteristics above and enforcing the iBDM selection criteria onto the small number of atmospheric neutrino events in the data sample leaves negligible levels of backgrounds for the present analysis.

\subsection{CNGS Beam Neutrino Background}
The ICARUS experiment collected CNGS beam neutrino interactions with a dedicated trigger system that utilized the sum of PMT signals together with the CNGS beam ``early warning'' signal of an imminent proton extraction from the Super Proton Synchrotron, 2 spills of $10.5~\mu$s time width, separated by 50~ms, every 6 seconds~\cite{ICARUSTrigger}.
A 60~$\mu$s width CNGS-gate signal was opened according to the predicted neutrino spill arrival, fully enveloping the 10.5~$\mu$s proton extraction time, enabling the full acquisition of CNGS neutrino interaction events.

The time synchronization between CERN and Gran Sasso had 1\% -- 4\% inefficiencies due to missing ``early warning'' messages, causing potential beam neutrino events incorrectly tagged as those recorded out of the $60~\mu$s readout window, such as the atmospheric neutrino events. These events, however, were recovered and correctly tagged as the CNGS neutrino beam events through an offline procedure that compares the event timestamp with the beam spill extraction time database. 

Since the total readout window for the CNGS neutrino beam was 120~$\mu$s, every 6 seconds, the total fractional loss of the acceptance of the iBDM data sample $2\times10^{-5}$ is neglgible. Given the negligible acceptance loss, we take the conservative approach and remove any events triggered within the CNGS beam data readout time window to fully eliminate the backgrounds from CNGS neutrino interactions.

\section{Results}~\label{Results}
\begin{figure}[t]
    \includegraphics[width=1.0\linewidth, height=1.0\linewidth]{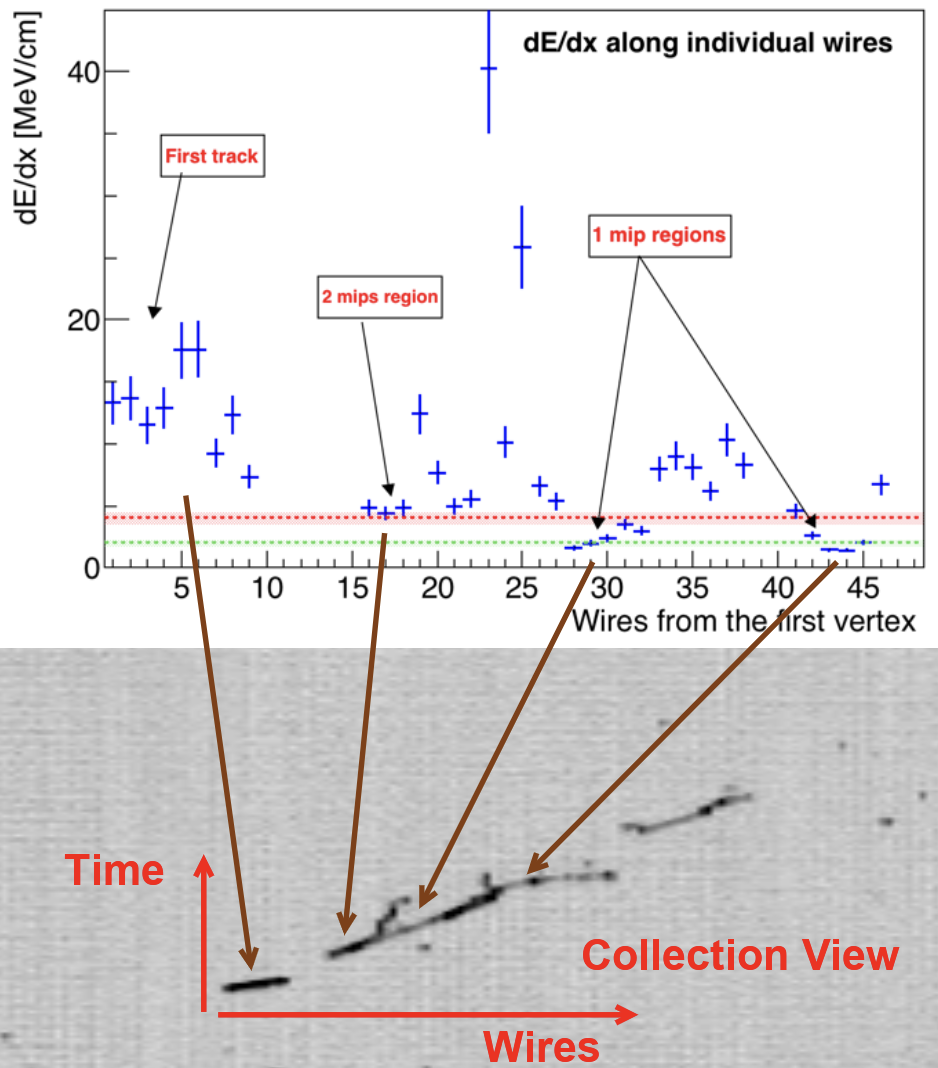}
    \hfill
    \caption{A zoomed-in Collection view of one of the final four selected iBDM candidates (bottom) and the corresponding $dE/dx$ energy deposit as a function of wires (top). The dimension of the event display image represents 40~cm (H)$\times$~50~cm (V) region of the detector. The brown arrows visually guide the wire numbers to the corresponding regions in the event display.  The time and wire information is the same as referenced in Fig.~\ref{fig:muoneventdisplay}. Based on the shower development pattern in the event display, the direction of the particle motion is from left to right. Topologically, two main interactions are recognizable in the event, with the left-most track resembling a primary interaction followed by the secondary one.}
    \label{fig:iBDMRejectedCandidate}
\end{figure}

\begin{figure}[htp!]
    \begin{center}
    \subfigure[]{\includegraphics[width=0.80\linewidth]{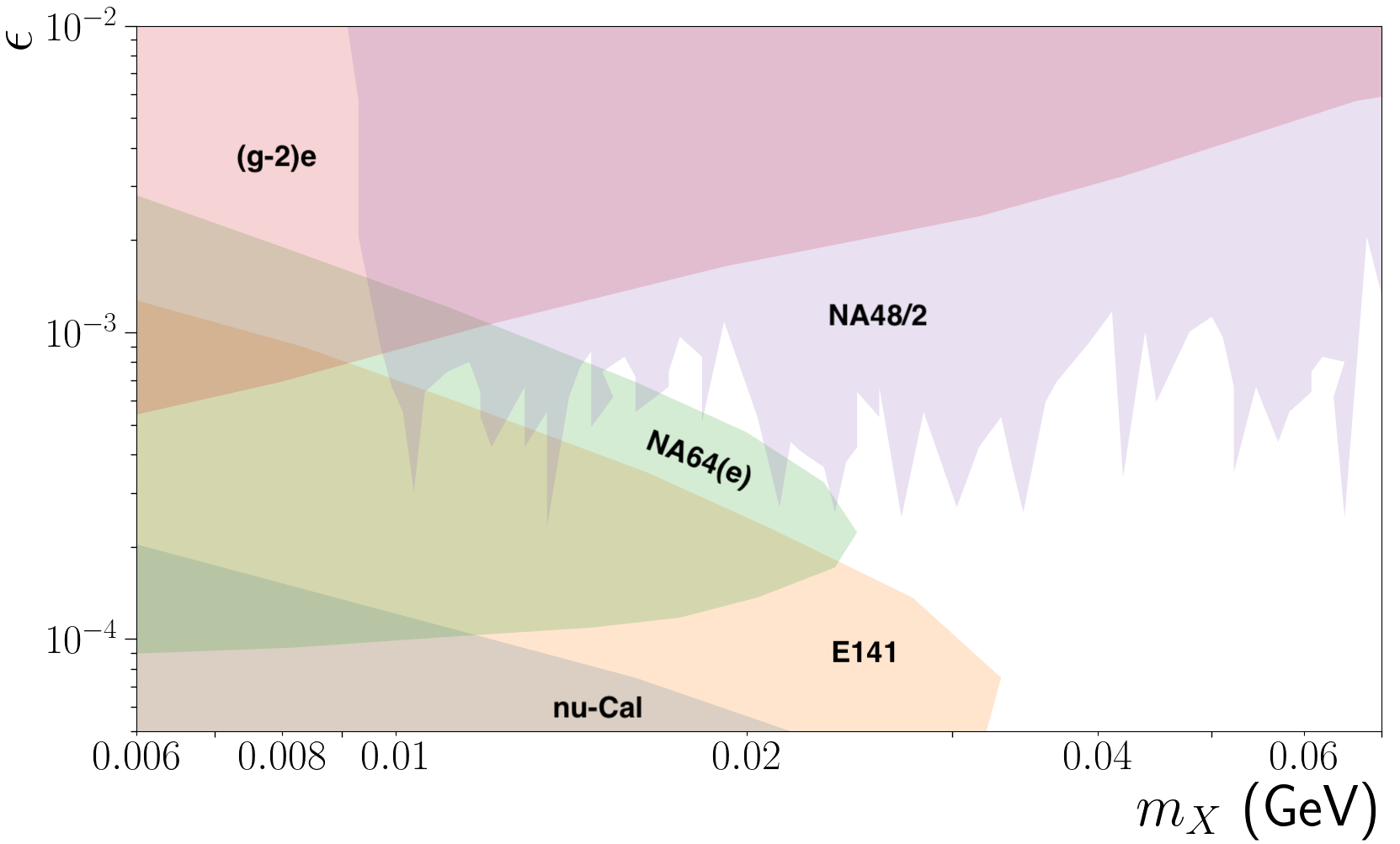}}
    \subfigure[]{\includegraphics[width=0.80\linewidth]{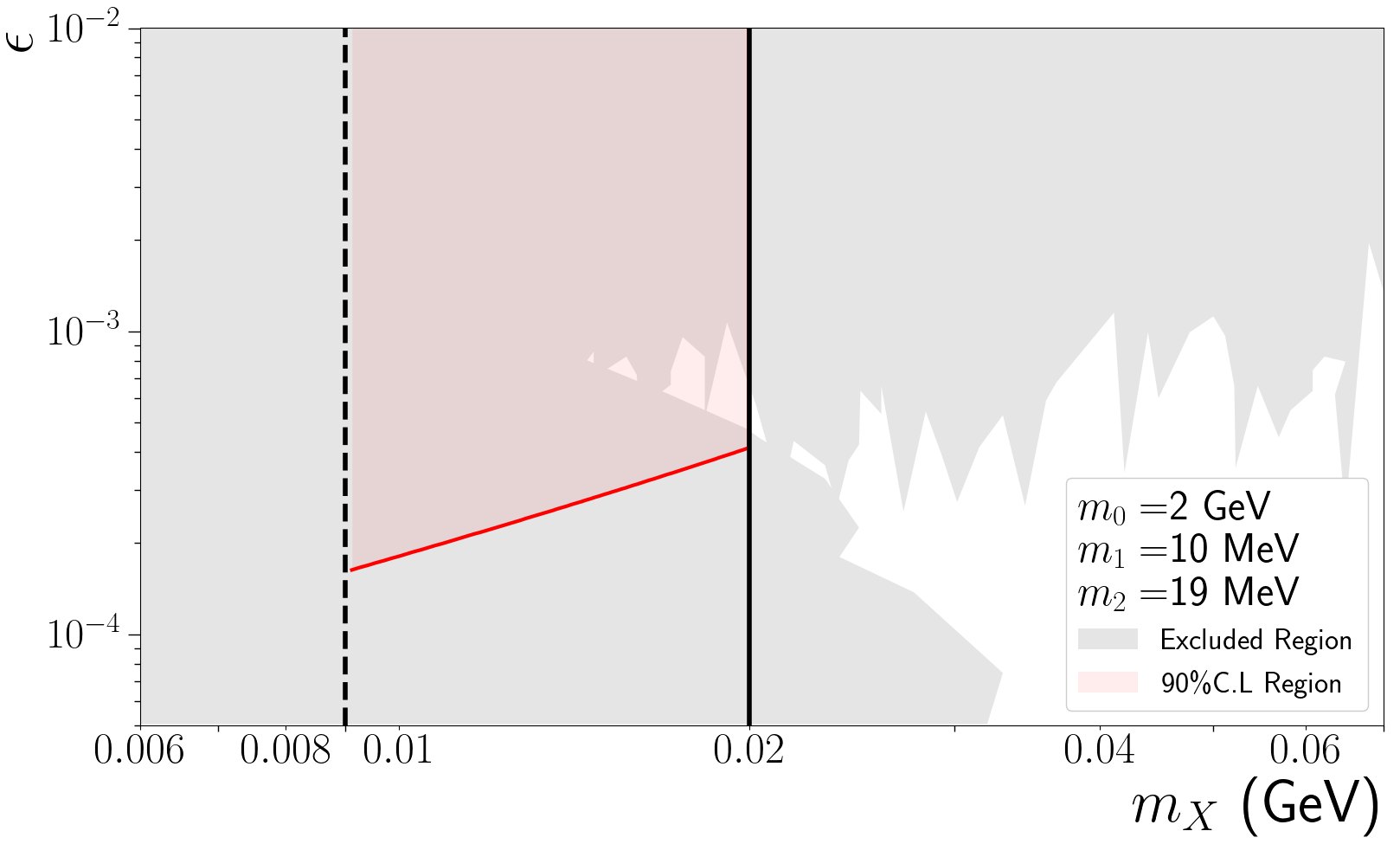}}
    \subfigure[]{\includegraphics[width=0.80\linewidth]{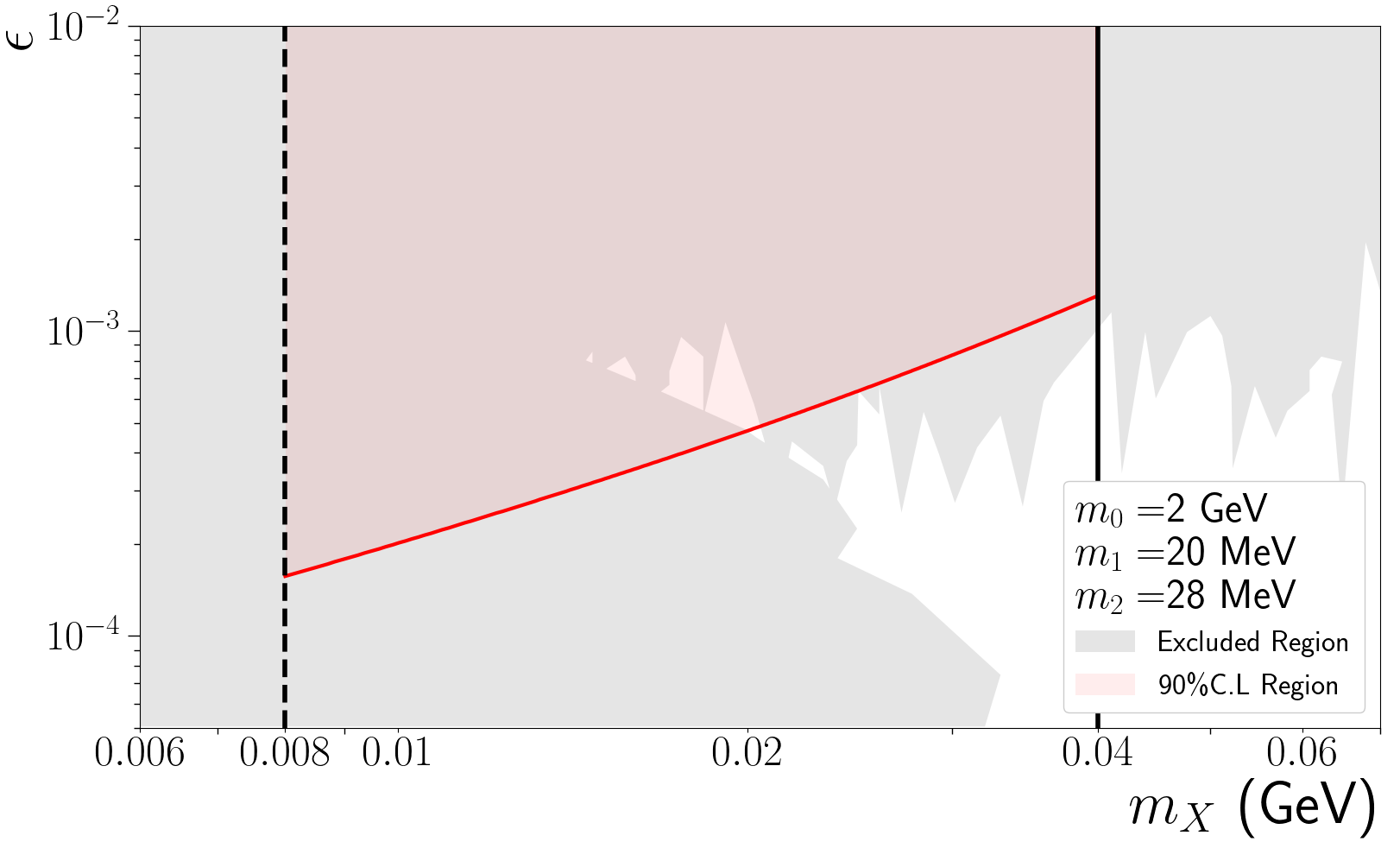}}
    \subfigure[]{\includegraphics[width=0.80\linewidth]{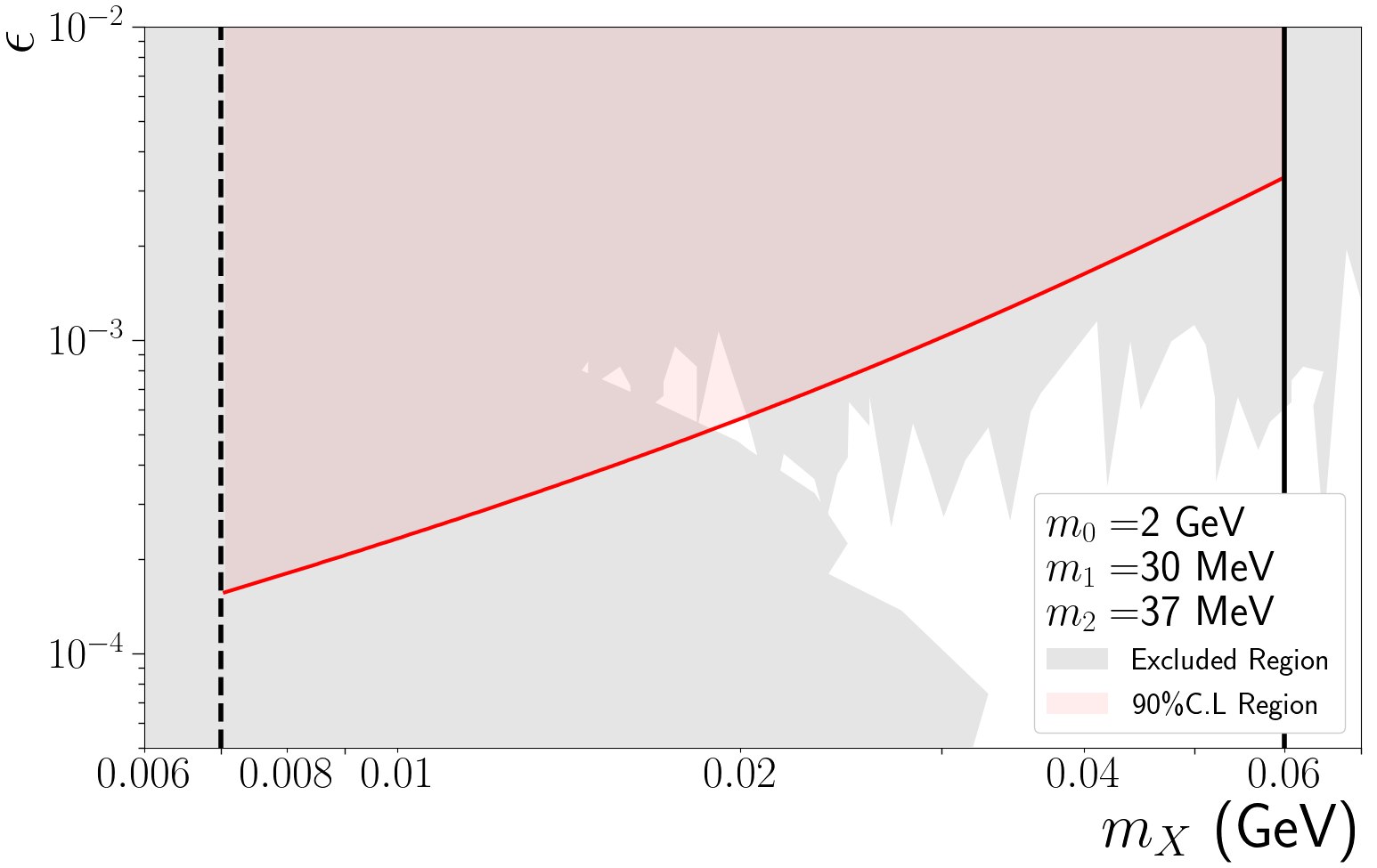}}
    \end{center}
    \caption{(a) Excluded $(m_X,\epsilon)$ parameter space by previous experiments, (g-2)e \cite{gminus2}, nu-Cal~\cite{nucal2011}, E141~\cite{E141}, NA64(e)~\cite{NA64e} and NA48/2~\cite{NA48}. (b) - (d) The corresponding 90 \%~C.L. limits (solid red line) and the excluded parameter region (red shade) set by this study for $m_0=2$~GeV and $(m_1, m_2) =$ (10~MeV,19~MeV), (20~MeV,~28~MeV), and (30~MeV,~37~MeV), respectively. The black lines show the kinematic limits on which the final states are either unobservable - solid line - or fail the minimum distance and the fiducial volume containment requirements.}\label{fig:exclusion}
\end{figure}

The preliminary scan identified more than 100 events with isolated showers some of these showers with a muon or completely isolated. After the background rejection process in Sec.~\Ref{Backgrounds}, four iBDM candidate events survived. These candidate events underwent a detailed, visual inspection of the $dE/dx$ of each shower to identify clear indications of one m.i.p. signature for the electron from the primary interaction and, subsequently, the two-m.i.p. signature for the associated $e^{+}e^{-}$ pair from the secondary interaction observed in the iBDM topology seen in the simulation. 
After thoroughly inspecting the $dE/dx$ characteristics of the showers in each of the four candidate events, three failed the $dE/dx$ requirements, while the fourth failed the fiducial volume containment requirement, resulting in zero observed events. 

To illustrate the final inspection, Figure~\ref{fig:iBDMRejectedCandidate} shows a zoomed-in view of the core of a rejected iBDM candidate event that failed the $dE/dx$ requirement. The bottom image shows the event display of the candidate in Collection view, while the $dE/dx$ energy deposit of the corresponding wires is shown on the top. 
The time and wire information is the same as referenced in Fig.~\ref{fig:muoneventdisplay}. 
\newline 
Topologically two interaction vertices are clearly recognizable in the event which proceeds from left to right, based on the direction of the shower development of the right-most track. The left-most, short track is associated with the primary interaction, while the right-most, showering track, is associated with the secondary interaction.
\newline
The top graph shows the $dE/dx$ as a function of wires from the start of the first track in the event display.
The light green horizontal dashed line indicates the energy deposition that corresponds to the 1 m.i.p. signal, while the red horizontal dashed line corresponds to 2 m.i.p. signal.
\newline
Inspecting the first few wires of the secondary interaction implies a 2 m.i.p. signal, which confirms the secondary interaction is an electron-positron pair that then proceeds to shower, consistent with the secondary interaction of an iBDM event.  
It has to be emphasized that the detector is capable of clearly identifying the 1 m.i.p signature of a minimum ionizing electron, such that the two separate m.i.p. signatures are recognized inside the shower in the secondary interaction.
\newline
However, the left-most short track of the primary interaction cannot be associated to a single 1 m.i.p. electron because of the much higher $dE/dx$ energy deposition at several m.i.p. level. Therefore, this event is rejected.

With the null observation, the 90\%~C.L. exclusion limits for the $(m_X,\epsilon)$ parameter space have been set slightly improving the existing limits for $m_X \sim 17 $ MeV and $\epsilon \sim 6 \times 10^{-4}$ for some $m_0, m_1, m_2$ explored parameters. 
Figure~\ref{fig:exclusion}.(a) shows the excluded regions achieved by beam dump experiments (e.g., nu-Cal \cite{nucal2011}, E141 \cite{E141}) and collider/fixed target experiments (e.g., NA64(e) \cite{NA64e}, NA48/2  \cite{NA48}) and Fig.~\ref{fig:exclusion}.(b) - (d) show the 90\% CL exclusion regions achieved by this study for the three optimal mass sets ($m_1,m_2$) = (10 MeV, 19 MeV), (20 MeV, 28 MeV), and (30 MeV, 37 MeV), respectively, for the reference case of $m_0=$~2~GeV identified in Sec.~\ref{Analysis}. 

Given the greater number of $N_{\rm expected}$ seen in Table~\ref{table:1}, and just the 12\% difference in the number of total events greater than the 200~MeV energy threshold set by the trigger, the exclusion plots for $m_0=$~1~GeV and ($m_1,m_2$) = (10 MeV, 18 MeV), (20 MeV, 26 MeV), and (30 MeV, 36 MeV) are similar, spanning more available parameter space than the mass sets for $m_0=$~2~GeV.
The solid black line in Fig.~\ref{fig:exclusion}.(b) -- (d) represents the dark photon mass limit ($m_X>2m_1$) above which the probability for visible final states in the detector is low. The dashed black line represents the dark photon mass limit ($m_X<m_2-m_1$) below which the secondary vertex is too close to be distinguished from the primary vertex.
The solid red line in each plot represents the central value of the 90\% CL exclusion limit with the light red shaded area indicating the parameter space excluded by our results. 
As can be seen in these plots, our results completely cover the hole  left between NA64(e)\cite{NA64e} and NA48/2~\cite{NA48} experiments.

\section{Conclusions}\label{Conclusions}

In this paper, we present an iBDM search using the data taken by the ICARUS LArTPC detector during its 2012 -- 2013 operation.  
The data set used for the search corresponds to a total exposure of 0.13 kton$\cdot$year and contains a total of 4,134 events that passed the atmospheric neutrino event filter which requires the presence of at least one e.m. shower. 
The iBDM signature sought in this analysis requires an electron from the primary interaction followed by an associated $e^{+}e^{-}$ pair.
The search results in zero observed events, and the exclusion limits are set in the dark photon ($m_X, \epsilon$) parameter space for ($m_0,m_1,m_2$) mass sets as shown in the exclusion plots in Fig.~\ref{fig:exclusion}.

Despite the rather small total exposure of data analyzed in this paper, the ICARUS Gran Sasso iBDM search expands the excluded parameter space beyond the previously explored region. 
This result leverages the precision 3D imaging and energy measurement capabilities of the LArTPC, as well as the large overburdened location of the experiment, which greatly reduces the background from cosmic rays.
In this regard, this result indicates an excellent opportunity for large-scale future neutrino experiments, such as DUNE to greatly expand the parameter space and potentially discover an iBDM.

\section*{Acknowledgments}
The work of HC and JY is supported by the U.S. Department of Energy under Grant No. DE-SC0011686.
The work of SB, BB, SC, and RR is supported partially by the University of Texas at Arlington and the U.S. Department of Energy under Grant No. DE-SC0011686.
The work of DK is supported partially by the University of South Dakota.
The fundamental support of INFN to ICARUS LArTPC program is warmly acknowledged.

\typeout{}
\bibliography{apssamp}

\end{document}